# Transient Stability Analysis of a Hybrid Grid-Forming and Grid-Following RES System Considering Multi-Mode Control Switching

Ruiyuan Zeng[1], *Student Member, IEEE*, Ruisheng Diao[1*], *Senior Member, IEEE*, Fangyuan Sun[1], *Student Member, IEEE*, Wangqianyun Tang[2,3], Junjie Li[2,3], Baorong Zhou[2,3]

***Abstract*—The inherent control switching of renewable energy sources (RESs) during intricate transient processes introduces complexity to the dynamic behavior of modern power systems. This paper reveals the dynamic coupling between grid-forming (GFM)/grid-following (GFL)-based RES and dominant instability modes of the hybrid system. First, six control combinations are systematically investigated by pairing the two GFM-RES modes, normal control and current saturation, with the three GFL-RES modes: normal control, low voltage ride-through (LVRT), and high voltage ride-through (HVRT). Based on switching system theory, the coupled power flow and dynamic motion models are developed considering multi-mode switching characteristics. It is revealed that the hybrid system exhibits two distinct instability modes when the GFM-RES and GFL-RES exceed their *P-f* and *V-f* desynchronization boundaries, respectively. The two-dimensional spatiotemporal damping characteristics of GFL-RES induced by GFM-RES are also uncovered for the first time. A novel criterion is proposed to quantify the impact of GFM-RES on GFL-RES dynamics, capturing both its stabilizing and destabilizing effects under different control combinations. High-fidelity electromagnetic transient simulations validate the correctness of the analysis framework.**

## Nomenclature

| | |
|---|---|
| RES | Renewable Energy Source |
| GFM/GFL | Grid-Forming/Grid-Following |
| LVRT/HVRT | Low/High Voltage Ride-Through |
| SEP/UEP | Stable/Unstable Equilibrium Point |

This work is supported by the National Natural Science Foundation of China (U22B6007) and Science and Technology Project of China Southern Power Grid Co. Ltd. (ZBKJXM20232419).

R. Zeng, R. Diao and F. Sun are with the [1]ZJU-UIUC Institute, Zhejiang University, Haining, 314400 China (e-mail: ruiyuan.23@intl.zju.edu.cn, fysun_zju.zju.edu.cn).

W. Tang, J. Li, and B. Zhou are with [2]China Southern Grid Electric Power Research Institute Co., Ltd, and [3]China Southern Power Grid Co., Ltd, Guangzhou, 510623 China.

Corresponding author: Prof. Ruisheng Diao, email: ruishengdiao@intl.zju.edu.cn

## I. Introduction

With the continued deepening of green energy policy, RES integration projects are being actively promoted worldwide based on direct current (DC) techniques. The ever-increasing penetration of RESs, together with the gradual retirement of conventional power plants and continuous growth of electricity demand, has led to a "hollowing-out" phenomenon of the power grid. As a result, voltage support in load centers has weakened, creating an urgent need for reinforcement. A hybrid integration of GFM and GFL- RESs provides a suitable solution to this challenge, where GFM control is primarily designed to provide voltage support to systems, while GFL control focuses on maximizing energy transfer efficiency.

The significant differences between RES and traditional fossil-fuel-based sources, including the fundamental operating principles and hardware characteristics, introduce unique challenges in control and dispatch that diverge substantially from the conventional generation units [1], [2]. For GFM converter-based-RES, it is prone to encounter current saturation issues during transient processes due to the weak overcurrent tolerance of power electronic components and the inherent voltage source characteristics. For GFL-RES, due to its voltage synchronization mechanism [3], it should comply with the LVRT/HVRT requirements, as mandated by national grid codes. Therefore, the control switching behaviors of GFM/GFL-RES have gained increasing attention in recent years for maintaining system stability and security following major disturbances.

In recent years, stability incidents triggered by the tripping RESs have occurred across various countries, including Odessa in the United States [4], South Australia [5], and parts of the United Kingdom [6]. The majority of these incidents undergo abnormal transient control switching or overcurrent issues, thereby causing growing concerns. The transient stability analysis considering the switching characteristics of GFM-RES has been investigated in [7]-[14]. Ref [7] investigates the instability mode of GFM-RES governed by current saturation. The reshaping of the power-angle relationship increases the vulnerability of GFM-RES to transient instability [7], [8]. A stability enhancement strategy is proposed to maintain the stable operation of GFM-RES during transients [8]. However, the absence of a desaturation mechanism renders it susceptible to continuous current saturation issues, which is mitigated in [9], [10] by employing anti-windup strategies. Refs [11], [12] demonstrate that the shrink of the GFM-RES's attraction region occurs after current saturation imposes further requirements on transient control design. Similar conclusions are drawn through critical clearing time analysis [13], [14]. Despite that, the GFL-RES also undergoes transient control switching events, its *Q-f* characteristic, which is dual to *P-f* synchronized devices, introduces heterogeneous dynamics into systems [1], [2]. Ref [15] investigates the dynamics of GFL-RES from the power synchronization perspectives while proposing a synchronization enhancement transient strategy. Focusing on the *K*-factor of the LVRT strategy, the existence of stable equilibrium point (SEP) is investigated in [16]. Ref [17] reveals the evolution of the attraction region of GFL-RES when considering the switching characteristic dominated by LVRT control. Furthermore, Ref [18] uncovers the mechanism of repeated LVRT issues based on switched system theory, which has also been observed in the northern grid and southern grid of China [19], [20].

However, all the aforementioned studies [7]-[18] are limited to the control switching characteristics of the single-machine-infinite-bus system while neglecting the energy diversity in practical modern power systems. The investigation of transient stability of multi-machine systems with multi-type RESs is still in its early stages, especially with regard to scenarios characterized by multi-mode switching dynamics. Considering the LVRT strategy, Ref [21] proposed a transient stability assessment method for multi-paralleled wind farms by solving the existence of SEPs of wind farms during fault. However, evaluating transient stability solely from the perspective of SEP existence yields results that are either conservative or overly optimistic. Ref [22] derives the dynamic interaction between GFM/GFL-RES under the assumption of fixed control modes. Ref [23] analyzes the dynamic model of systems with GFM/GFL-RES and synchronous generator (SG), while systematically proposing a nonlinear backstepping control. Nevertheless, it lacks in-depth investigation into stability mechanisms and provides limited analysis of the impact of transient control switching. Ref [24] conducted a modeling study on a grid-connected system with parallel GFM-RES, examining the interaction characteristics of two machines considering current saturation control switching. However, it lacks an in-depth theoretical analysis. Ref [25] reveals the mechanical-electrical coupling between GFM/GFL-RES and illustrates a new electromechanical synchronous issue. Still, only the transient switching characteristics of GFL-RES are considered. Considering the dynamic characteristics of low-inertia power systems, Ref [26] applied the Zubov method to address the challenge of attraction region estimation in the presence of negative damping introduced by GFL-RES. Its theoretical model is also equivalent to a two-machine system. Nonetheless, the studied two-machine system in [25] and [26] can be characterized by one independent relative angle, which simplifies the multi-machine interaction. To summarize, most existing efforts primarily focus on control switching governed by a single source, neglecting the coupling of multi-mode switching phenomena that arise from interactions among multiple machines.

This paper investigates the dynamic mechanisms of GFM/GFL-RES hybrid systems under multi-mode control switching, establishing six control combination regions to define switching boundaries and operating modes. It reveals energy-based interactions and two distinct instability modes. Moreover, a unique two-dimensional damping characteristic is identified owing to GFM/GFL-RES interactions. The main contributions of this paper are summarized as follows:

1) The feasible regions corresponding to the six control combinations of the GFM/GFL-RES hybrid systems are uncovered for the first time, i.e., resulting from the pairing of two GFM-RES modes (normal control and current saturation) and three GFL-RES modes (normal control, LVRT, and HVRT).

2) The energy transformation and dissipation of the hybrid systems are derived considering the coupling of power flow coupling and dynamics, in which the former influences the potential force of the GFM/GFL-RSE, and both jointly affect the damping characteristics of GFL-RES.

3) Based on the theoretical analysis, the principles for identifying the first-swing instability modes of the GFM/GFL-RES hybrid system are proposed under critical conditions. Meanwhile, the criterion for determining the GFM-RES's effect on GFL-RES is also developed within the context of the identified two-dimensional damping of GFL-RES.

The remainder of the paper is organized as follows. Section II establishes the mathematic models for both GFM/GFL-RES and the hybrid system. Section III derives the feasible regions of six control combinations and further investigates the energy transformation and dissipation characteristics of the system. Based on theoretical analysis, the first-swing instability is identified, along with the mechanisms by which GFM-RES may either enhance or deteriorate the stability of GFL-RES. Section IV validates the effectiveness of the proposed multi-machine analysis framework. Section V draws the conclusions.

## II. Modeling of GFM/GFL-RES and the System

This section introduces the basic model of switching system theory. The modeling at both the device and system levels is illustrated considering control switching.

### *A. Switched System Theory*

A continuous-time dynamical system with discrete switching characteristics [27] can be described as:

$$\begin{cases} \dot{\boldsymbol{x}} = \boldsymbol{f}_n(\boldsymbol{x}) \\ \boldsymbol{y} = \boldsymbol{g}_n(\boldsymbol{x}) \\ n = \sigma(\boldsymbol{y}) \end{cases} \tag{1}$$

where $\boldsymbol{x}$ denotes the state vector, which determines the future dynamic evolution of the system. $\boldsymbol{y}$ denotes the output vector and affects the switching process. $n$ denotes $n^{\text{th}}$ subsystem. $\boldsymbol{f}_n(\cdot)$ and $\boldsymbol{g}_n(\cdot)$ indicate the dynamic function and output function, respectively. $\sigma(\cdot)$ denotes the switching function that governs the switching mechanism.

In this paper, the GFM/GFL-RES hybrid system is a multi-input multi-output system, where $\boldsymbol{y} \in \mathbb{R}^2$ consists of: 1) $I_{\text{FM,dq}}^{\text{ref}}$, the current magnitude of the inner voltage loop output of GFM-RES, and 2) $U_{\text{FL}}$, the voltage magnitude of GFL-RES at the point of common coupling (PCC). For clarity and readability, the subsequent analysis adopts the following simplifications: the relative virtual power angles $\delta_{12}$ and $\delta_{13}$ are used to denote the elements of $\boldsymbol{x}$; the synchronization expressions of PSL and PLL are used to formulate $\boldsymbol{f}_n(\cdot)$; $I_{\text{FM,dq}}^{\text{ref}}$ and $U_{\text{FL}}$ together with their corresponding expressions are used to represent the elements of $\boldsymbol{y}$ and $\boldsymbol{g}_n(\cdot)$; and the subsystem index is directly denoted by $n$ instead of $\sigma(\boldsymbol{y})$.

Since GFM-RES includes two control modes (normal control and current saturation) and the GFL-RES includes three modes (LVRT, normal control, and HVRT), the total number of control combinations is six, i.e., $n=6$. $n=1\sim3$ and $n=4\sim6$ represent the control combinations where the GFM-RES operates in normal control and current saturation modes, respectively, while the GFL-RES operates under the three aforementioned controls.

### *B. Dynamic Model of GFM/GFL-RES*

*1) GFM-RES*: The commonly used control scheme of GFM-RES is shown in Fig. 1.

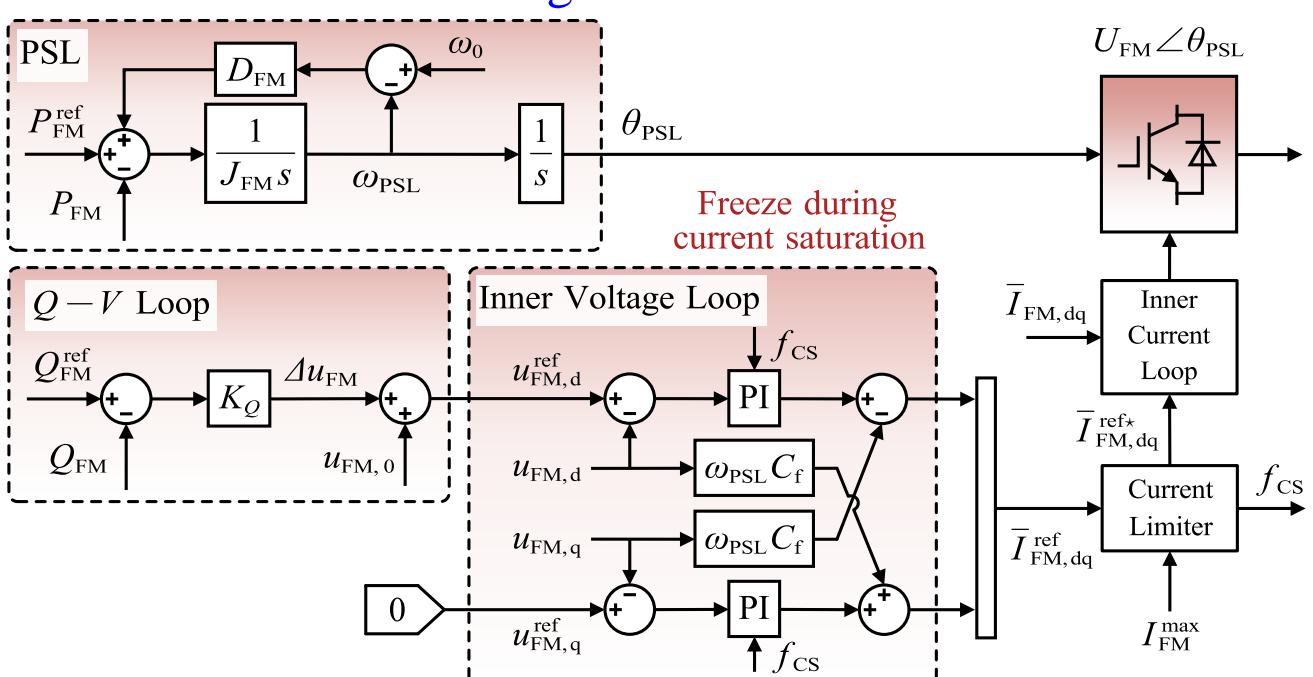

Fig. 1. Control scheme of GFM-RES with virtual synchronous generator.

The power synchronous loop (PSL) and *Q-V* loop enable the converter to provide basic frequency support and voltage regulation for the system. During the transient period, the current limiter outputs the preset reference $\bar{I}_{\text{FM,dq}}^{\text{sa}}$ to prevent overcurrent issues when $I_{\text{FM,dq}}^{\text{ref}}$ exceeds the threshold. At the same time, the current saturation flag steps from 0 to 1, which freezes the integrator in the inner voltage loop, preventing abnormal accumulation. This "freezing strategy" is widely employed in engineering. According to Fig. 1, the output function of GFM-RES can be expressed as:

$$I_{\text{FM,dq}}^{\text{ref}} = \sqrt{\begin{array}{l}[K_{\text{pd}}^{\text{IV}}(u_{\text{FM,d}}^{\text{ref}} - u_{\text{FM,d}}) + x_{\text{d}}^{\text{IV}}]^2 + \\ {[K_{\text{pq}}^{\text{IV}}(u_{\text{FM,q}}^{\text{ref}} - u_{\text{FM,q}}) + x_{\text{q}}^{\text{IV}}]^2}\end{array}} \tag{2}$$

$$u_{\text{FM,d}}^{\text{ref}} = K_Q(Q_{\text{FM}}^{\text{ref}} - Q_{\text{FM}}) + u_{\text{FM,0}} \tag{3}$$

*2) GFL-RES*: Fig. 2 presents the transient control scheme of GFL-RES, including LVRT and HVRT controls. The basic principles of this transient strategy are to increase or decrease the reactive power injection during low-voltage or high-voltage conditions, respectively, thereby preventing RESs from tripping and accelerating post-fault voltage recovery. Hence, this voltage-based control naturally divides the control range into three segments: $(-\infty, U_{\text{FL}}^{\text{LV}})$, $[U_{\text{FL}}^{\text{LV}}, U_{\text{FL}}^{\text{HV}}]$, and $(U_{\text{FL}}^{\text{HV}}, +\infty)$. When $U_{\text{FL}} \in [U_{\text{FL}}^{\text{LV}}, U_{\text{FL}}^{\text{HV}}]$, the GFL-RES operates under normal control, i.e.,

$$\begin{cases} I_{\text{FL}} = I_{\text{FL,0}} \\ \varphi_{\text{FL}} = \varphi_{\text{FL,0}} \end{cases} \tag{4}$$

When $U_{\text{FL}} \in (-\infty, U_{\text{FL}}^{\text{LV}})$, the GFL-RES operates under LVRT control, satisfying

$$\begin{cases} I_{\text{FL}} = \min[K_I^{\text{RT}}(U_{\text{FL}}^{\text{LV}} - U_{\text{FL}}) + I_{\text{FL,0}}, I_{\text{FL}}^{\max}] \\ \varphi_{\text{FL}} = \max\left[K_\varphi^{\text{RT}}(U_{\text{FL}} - U_{\text{FL}}^{\text{LV}}) + \varphi_{\text{FL,0}}, -\dfrac{\pi}{2}\right] \end{cases} \tag{5}$$

If $U_{\text{FL}} \in (U_{\text{FL}}^{\text{HV}}, +\infty)$, the HVRT control is triggered, i.e.,

$$\begin{cases} I_{\text{FL}} = \min[K_I^{\text{RT}}(U_{\text{FL}} - U_{\text{FL}}^{\text{HV}}) + I_{\text{FL,0}}, I_{\text{FL}}^{\max}] \\ \varphi_{\text{FL}} = \min\left[K_\varphi^{\text{RT}}(U_{\text{FL}} - U_{\text{FL}}^{\text{HV}}) + \varphi_{\text{FL,0}}, \dfrac{\pi}{2}\right] \end{cases} \tag{6}$$

Since there is no fundamental difference among GFL-RESs' LVRT/HVRT grid code across countries [17], this paper adopts the Chinese grid code as the reference [17], by setting $[U_{\text{FL}}^{\text{LV}}, U_{\text{FL}}^{\text{HV}}] = [0.9, 1.1]$.

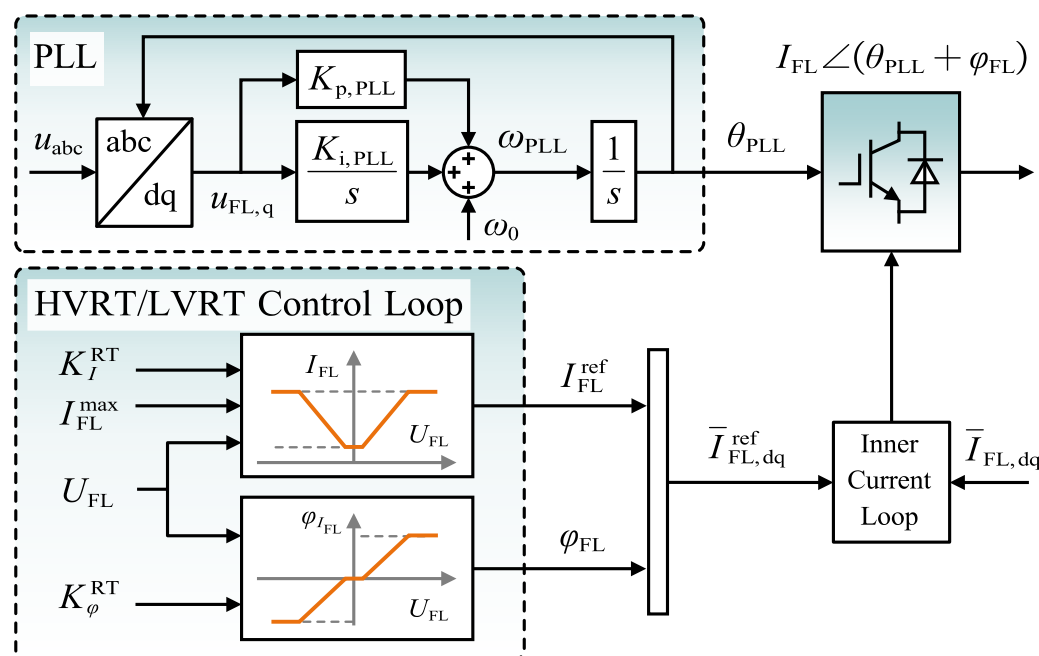

Fig. 2. Control scheme of GFL-RES with FRT control.

### *C. Power Flow Model of the Hybrid Power System*

The typical topology of the GFM/GFL-RES hybrid system is given in Fig. 3, where the main grid is modeled as an ideal voltage source to simplify the subsequent muti-machine interaction analysis. Bus 1 to 3 are active nodes, and bus 4 is passive node. The algebraic admittance model can be described as:

$$\begin{bmatrix} \mathbf{I}_{\text{s}} \\ 0 \end{bmatrix} = \begin{bmatrix} \mathbf{Y}_{\text{a}} & \mathbf{Y}_{\text{b}} \\ \mathbf{Y}_{\text{c}} & \mathbf{Y}_{\text{d}} \end{bmatrix} \begin{bmatrix} \mathbf{U}_{\text{s}} \\ \mathbf{U}_{\text{ns}} \end{bmatrix} \tag{7}$$

where subscript "s" and "ns" indicate active node and passive node, respectively. $\mathbf{U}$ and $\mathbf{I}$ denote voltage vector and current vector. $\mathbf{Y}_{\text{a}}$, $\mathbf{Y}_{\text{b}}$, $\mathbf{Y}_{\text{c}}$, and $\mathbf{Y}_{\text{d}}$ are corresponding sub-matrices of the admittance matrix, which are also affected by system fault and load changes. Employing the Kron-reduction method to eliminate the passive nodes, (7) can be rewritten as:

$$\mathbf{I}_{\mathrm{s}}=(\mathbf{Y}_{\mathrm{a}}-\mathbf{Y}_{\mathrm{b}}\mathbf{Y}_{\mathrm{d}}^{-1}\mathbf{Y}_{\mathrm{c}})\mathbf{U}_{\mathrm{s}}=\mathbf{Y}_{\mathrm{r}}\mathbf{U}_{\mathrm{s}} \tag{8}$$

The expanded form of (8) is:

$$\begin{bmatrix}\mathbf{I}_{\mathrm{V}}\\ \mathbf{I}_{\mathrm{I}}\end{bmatrix}=\begin{bmatrix}\mathbf{Y}_{\mathrm{ra}} & \mathbf{Y}_{\mathrm{rb}}\\ \mathbf{Y}_{\mathrm{rd}} & \mathbf{Y}_{\mathrm{rc}}\end{bmatrix}\begin{bmatrix}\mathbf{U}_{\mathrm{V}}\\ \mathbf{U}_{\mathrm{I}}\end{bmatrix} \tag{9}$$

where subscript "V" and "I" indicate the voltage source and current source. To solve for the unknown variables, a matrix transformation is applied to move the current vector of current source nodes to the left-hand side of $\mathbf{Y}_{\mathrm{r}}$, i.e.,

$$\begin{bmatrix}\mathbf{I}_{\mathrm{V}}\\ \mathbf{U}_{\mathrm{I}}\end{bmatrix}=\mathbf{M}\begin{bmatrix}\mathbf{U}_{\mathrm{V}}\\ \mathbf{I}_{\mathrm{I}}\end{bmatrix} \tag{10}$$

$$\mathbf{M}=\begin{bmatrix}\mathbf{Y}_{\mathrm{ra}}-\mathbf{Y}_{\mathrm{rb}}\mathbf{Y}_{\mathrm{rd}}^{-1}\mathbf{Y}_{\mathrm{rc}} & \mathbf{Y}_{\mathrm{rb}}\mathbf{Y}_{\mathrm{rd}}^{-1}\\ -\mathbf{Y}_{\mathrm{rd}}^{-1}\mathbf{Y}_{\mathrm{rc}} & \mathbf{Y}_{\mathrm{rd}}^{-1}\end{bmatrix} \tag{11}$$

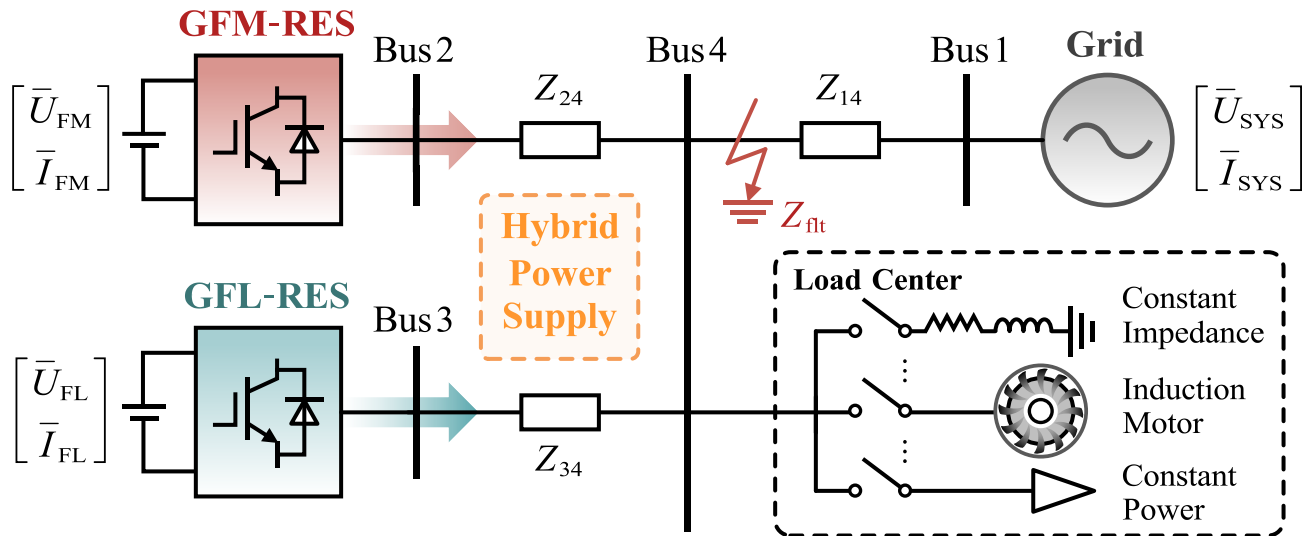

Fig. 3. Typical topology of hybrid power system with GFM/GFL-RESs.

By incorporating the dynamic characteristics of GFM/GFL-RES and the power flow properties of the system, the overall control interaction characteristics of the hybrid system can be obtained, as shown in Fig. 4.

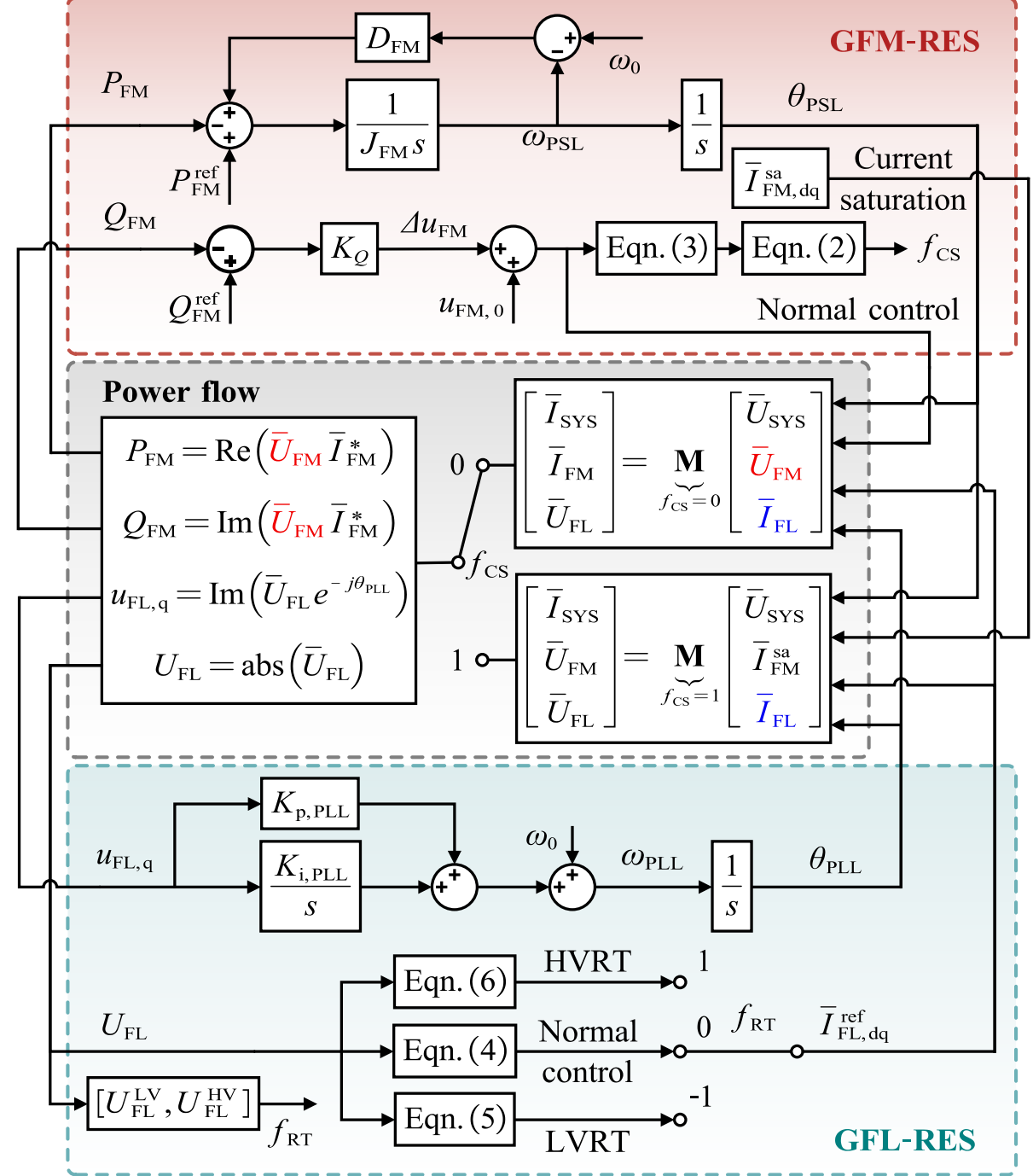

Fig. 4. Control interaction characteristics of GFM/GFL-RES hybrid system.

It should be noted that the phasor model is valid as long as the condition $\omega \approx \omega_0$ is satisfied, and this condition holds for the subsequent analysis in this paper. However, when analyzing extreme phenomena such as frequency collapse, it is necessary to adjust the analytical framework or adopt alternative methods for more detailed analysis.

## III. Multi-Mode Control Switching and Dynamic Coupling Characteristics

This section explores the control combination regions considering multi-mode control switching, while deriving the mechanism of energy transformation and dissipation of the hybrid system. Furthermore, the dominant instability principle under critical conditions is revealed. A novel criterion is proposed to determine whether the GFM-RES enhances or deteriorates the stability of GFL-RES.

### A. Multiple Control Combination Region

The control combination determines the dynamics of the device and the overall characteristics of the system. Hence, the primary target is to derive the feasible operating range $R_n$ and its boundary for the $n^{\mathrm{th}}$ control combination, $n=1\sim6$. Since the bandwidths of the inner voltage and current loops are much higher than that of the synchronization loop, the dynamics of the inner control loop can be neglected when analyzing transient stability issues caused by the synchronization loop, which is a common practice in large-disturbance studies [7]-[11], [13], [17]. The subsequent derivations are based on this assumption.

*1) GFM-RES Operates under Normal Control Mode (i.e., $n=1\sim3$)*: Once these control combinations are triggered, the relationship between $U_{\mathrm{FM}}$, $I_{\mathrm{FL}}$, and $\varphi_{\mathrm{FL}}$ should be solved firstly before calculating the whole system state variables. According to Fig. 1 and (3), *Q-V* loop couples the reactive power $Q_{\mathrm{FM}}$ and voltage magnitude $U_{\mathrm{FM}}$ of GFM-RES. Meanwhile, as the current magnitude $I_{\mathrm{FL}}$ and angle $\varphi_{\mathrm{FL}}$ of GFL-RES are determined by its volage of PCC $U_{\mathrm{FL}}$, they are consequently affected by $U_{\mathrm{FM}}$, as shown in Fig. 4 and (4)~(6). Under steady-state conditions, $u_{\mathrm{FM,d}}^{\mathrm{ref}}$ equals to $U_{\mathrm{FM}}$. By substituting $Q_{\mathrm{FM}}=\mathrm{Im}\left(\bar{U}_{\mathrm{FM}}\bar{I}_{\mathrm{FM}}^{*}\right)$ into (3), the function of $U_{\mathrm{FM}}$ can be derived in terms of $I_{\mathrm{FL}}$ and $\varphi_{\mathrm{FL}}$,

$$f_{U_{\mathrm{FM}}}(I_{\mathrm{FL}},\varphi_{\mathrm{FL}})=\frac{-a+\sqrt{a^2-4b}}{2} \tag{12}$$

$$a=\frac{\left[\begin{array}{c}M_{2,1}U_{\mathrm{SYS}}\sin\left(\delta_{12}-\varphi_{M_{2,1}}\right)+\dfrac{1}{K_Q}\\ M_{2,3}I_{\mathrm{FL}}\sin\left(\delta_{12}-\delta_{13}-\varphi_{M_{2,3}}-\varphi_{\mathrm{FL}}\right)\end{array}\right]}{M_{2,2}\sin\left(-\varphi_{M_{2,2}}\right)} \tag{13}$$

$$b=\frac{-(K_Q Q_{\mathrm{FM}}^{\mathrm{ref}}+u_{\mathrm{FM},0})}{K_Q M_{2,2}\sin\left(-\varphi_{M_{2,2}}\right)} \tag{14}$$

where $a$ and $b$ are intermediate variables. $\delta_{i,j}$ indicates the virtual angle difference of bus $j$ relative to bus $i$. $M_{i,j}$ and $\varphi_{M_{i,j}}$ denote the magnitude and phase angle of the elements in the $i^{\text{th}}$ row and $j^{\text{th}}$ column of $\mathbf{M}$.

At this point, $U_{\text{FL}}$ can be solved using the power flow equation in Fig. 4. Substituting $U_{\text{FL}}$ into (4)~(6) yields functions for $I_{\text{FL}}$ and $\varphi_{\text{FL}}$ (i.e., $f_{I_{\text{FL}}}$ and $f_{\varphi_{\text{FL}}}$). By simultaneously solving $f_{U_{\text{FM}}}$, $f_{I_{\text{FL}}}$, and $f_{\varphi_{\text{FL}}}$, the implicit relationship between $U_{\text{FM}}$, $I_{\text{FL}}$, and $\varphi_{\text{FL}}$ can be obtained, i.e.,

$$\begin{cases} U_{\text{FM}} - f_{U_{\text{FM}}}(I_{\text{FL}}, \varphi_{\text{FL}}) = 0 \\ I_{\text{FL}} - f_{I_{\text{FL}}}(U_{\text{FM}}, I_{\text{FL}}, \varphi_{\text{FL}}) = 0 \\ \varphi_{\text{FL}} - f_{\varphi_{\text{FL}}}(U_{\text{FM}}, I_{\text{FL}}, \varphi_{\text{FL}}) = 0 \end{cases} \tag{15}$$

*2) GFM-RES Operates under Current Saturation Mode (i.e., $n = 4\sim6$ )*: Due to the fact that GFM-RES switches to the current source during current saturation mode, $U_{\text{FM}}$ no longer participates in the closed-loop control directly, as shown in Fig. 4. Consequently, the system's state is only implicitly determined by $I_{\text{FL}}$ and $\varphi_{\text{FL}}$:

$$\begin{cases} I_{\text{FL}} - f_{I_{\text{FL}}}(I_{\text{FL}}, \varphi_{\text{FL}}) = 0 \\ \varphi_{\text{FL}} - f_{\varphi_{\text{FL}}}(I_{\text{FL}}, \varphi_{\text{FL}}) = 0 \end{cases} \tag{16}$$

*3) Distribution of Control Combination*: Following the derivation of the key state variables of the system, the feasible operation region for each control combination can be solved. The specific conditions for all combinations are summarized in TABLE I.

TABLE I
MULTI-MODE CONTROL SWITCHING CONDITION

| **n** | **GFM-RES's condition** | **GFL-RES's condition** |
|---|---|---|
| 1 | Normal control, $I_{\text{FM,dq}}^{\text{ref}} \in [0, I_{\text{FM}}^{\max}]$ | LVRT, $U_{\text{FL}} \in (-\infty, U_{\text{FL}}^{\text{LV}})$ |
| 2 | Normal control, $I_{\text{FM,dq}}^{\text{ref}} \in [0, I_{\text{FM}}^{\max}]$ | Normal control, $U_{\text{FL}} \in [U_{\text{FL}}^{\text{LV}}, U_{\text{FL}}^{\text{HV}}]$ |
| 3 | Normal control, $I_{\text{FM,dq}}^{\text{ref}} \in [0, I_{\text{FM}}^{\max}]$ | HVRT, $U_{\text{FL}} \in (U_{\text{FL}}^{\text{HV}}, +\infty)$ |
| 4 | Current saturation, $I_{\text{FM,dq}}^{\text{ref}} \in (I_{\text{FM}}^{\max}, +\infty)$ | LVRT, $U_{\text{FL}} \in (-\infty, U_{\text{FL}}^{\text{LV}})$ |
| 5 | Current saturation, $I_{\text{FM,dq}}^{\text{ref}} \in (I_{\text{FM}}^{\max}, +\infty)$ | Normal control, $U_{\text{FL}} \in [U_{\text{FL}}^{\text{LV}}, U_{\text{FL}}^{\text{HV}}]$ |
| 6 | Current saturation, $I_{\text{FM,dq}}^{\text{ref}} \in (I_{\text{FM}}^{\max}, +\infty)$ | HVRT, $U_{\text{FL}} \in (U_{\text{FL}}^{\text{HV}}, +\infty)$ |

As shown in Fig. 5, the blue and red areas indicate that GFM-RES operates in normal control and current saturation modes, respectively. Each color is shown in three shades corresponding to the voltage levels of GFL-RES. Lighter shades indicate lower voltages, with the lightest shade representing that GFL-RES operates in LVRT mode. From Fig. 5, it can be seen that the switching boundary between the normal control mode and the current saturation mode of the GFM-RES is approximately parallel to $\delta_{13}$ axis. This indicates that the control switching behavior of the GFM-RES is primarily influenced by $\delta_{12}$. In contrast, the control switching of the GFL-RES exhibits an outward-spreading pattern from a central point, suggesting that its control switching characteristic is largely governed by both $\delta_{12}$ and $\delta_{13}$. A detailed sensitivity analysis related to Fig. 5 is provided in Appendix A.

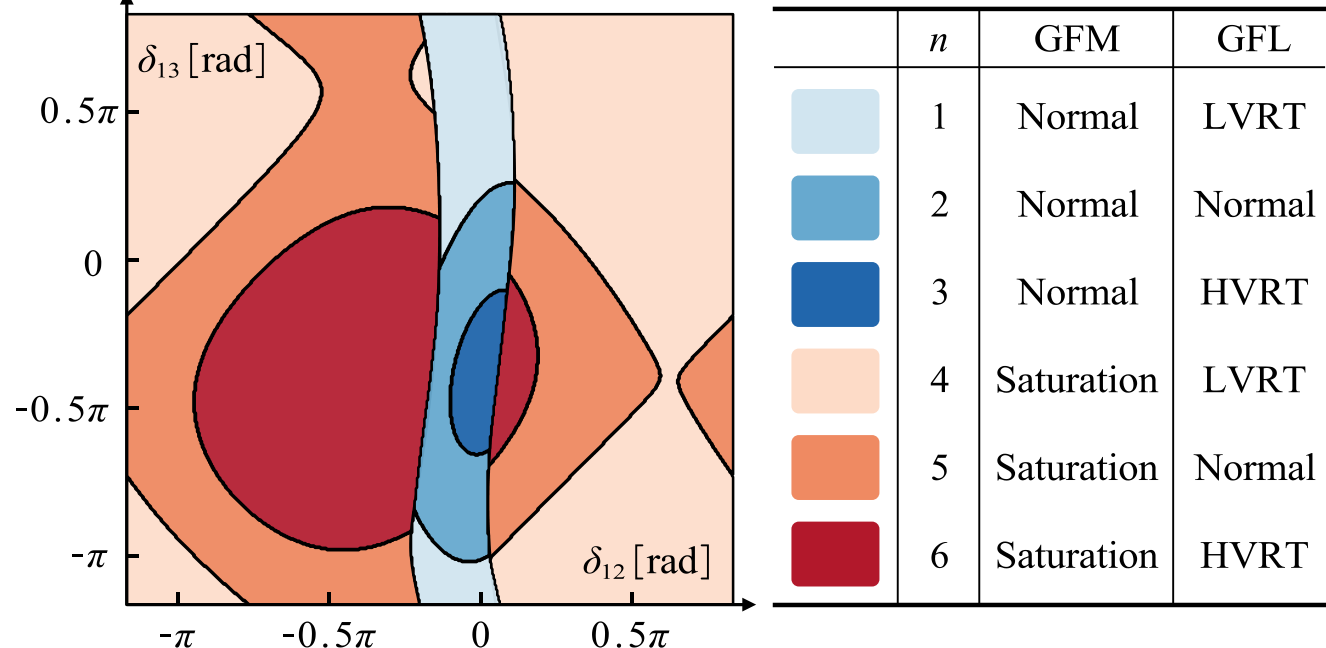

| n | GFM | GFL |
|---|---|---|
| 1 | Normal | LVRT |
| 2 | Normal | Normal |
| 3 | Normal | HVRT |
| 4 | Saturation | LVRT |
| 5 | Saturation | Normal |
| 6 | Saturation | HVRT |

Fig. 5. Distribution of control combination. The blue region indicates that the GFM-RES operates in the normal control mode, while the red region corresponds to the current saturation mode. The color shade represents different GFL-RES modes: the lightest shade denotes the LVRT mode, the medium shade the normal control mode, and the darkest shade the HVRT mode. In total, six control combinations are formed, as summarized in the table within the figure.

### *B. Energy Transformation and Dissipation under Multi-Mode Control Switching*

After deriving the multiple control combination regions in Section III.A, the corresponding system dynamics under each individual combination should be further analyzed.

*1) GFM-RES*: The dynamical synchronization process of GFM-RES is determined by PSL, as shown in Fig. 1. Based on Fig. 4, (10), and (11), the dynamic equation of GFM-RES can be expressed as follows:

$$J_{\text{FM}}\ddot{\delta}_{12} = \underbrace{P_{\text{FM}}^{\text{ref}} - P_{\text{FM}}}_{-F_{\text{FM,p}}} \underbrace{- D_{\text{FM}}\dot{\delta}_{12}}_{F_{\text{FM,d}}} \tag{17}$$

$$\underbrace{P_{\text{FM}}}_{n=1\sim3} = \begin{bmatrix} U_{\text{FM}} M_{2,1} U_{\text{SYS}} \cos\left(\delta_{12} - \varphi_{M_{2,1}}\right) + \\ (U_{\text{FM}})^2 M_{2,2} \cos\left(-\varphi_{M_{2,2}}\right) + \\ U_{\text{FM}} M_{2,3} I_{\text{FL}} \cos\left(\delta_{12} - \delta_{13} - \varphi_{M_{2,3}} - \varphi_{\text{FL}}\right) \end{bmatrix} \tag{18}$$

$$\underbrace{P_{\text{FM}}}_{n=4\sim6} = \begin{bmatrix} M_{2,1} U_{\text{SYS}} I_{\text{FM}}^{\text{sa}} \cos\left(\varphi_{M_{2,1}} - \varphi_{\text{FM}}^{\text{sa}} - \delta_{12}\right) + \\ M_{2,2} (I_{\text{FM}}^{\text{sa}})^2 \cos\left(\varphi_{M_{2,2}}\right) + \\ M_{2,3} I_{\text{FL}} I_{\text{FM}}^{\text{sa}} \cos\begin{pmatrix} \varphi_{M_{2,3}} - \varphi_{\text{FM}}^{\text{sa}} + \varphi_{\text{FL}} + \\ \delta_{13} - \delta_{12} \end{pmatrix} \end{bmatrix} \tag{19}$$

where $F_{\text{FM,p}}$ and $F_{\text{FM,d}}$ indicate the potential force and damping force of GFM-RES, respectively. $U_{\text{FM}}$, $I_{\text{FL}}$, and $\varphi_{\text{FL}}$ satisfy the implicit relationship solved in (15) and (16), which also hold in the subsequent derivations and will not be repeated.

Multiplying both sides of (17) by $\mathrm{d}\dot{\delta}_{12}/\mathrm{d}t$ and integrating, the dynamic equation of GFM-RES can be reformulated as $E_{\text{FM,k}} = E_{\text{FM,p}} + E_{\text{FM,d}}$, i.e.,

$$E_{\mathrm{FM,k}}=\frac{1}{2}J_{\mathrm{FM}}\left(\dot{\delta}_{12}\right)^{2} \tag{20}$$

$$E_{\mathrm{FM,p}}=\int_{C}-(P_{\mathrm{FM}}^{\mathrm{ref}}-P_{\mathrm{FM}})\mathrm{d}\zeta \tag{21}$$

$$E_{\mathrm{FM,d}}=\int_{C}\left(-D_{\mathrm{FM}}\dot{\delta}_{12}\right)\mathrm{d}\zeta \tag{22}$$

where $E_{\mathrm{FM,k}}$, $E_{\mathrm{FM,p}}$, and $E_{\mathrm{FM,d}}$ denotes the kinetic energy, potential energy, and energy dissipation of GFM-RES, respectively. $C$ indicates the path of the line integral, and $\mathrm{d}\zeta$ indicates a differential arc length element along the path.

Since the potential force of GFM-RES is associated with both $\delta_{12}$ and $\delta_{13}$, it can be regarded as a three-dimensional surface over the $\delta_{12}$-$\delta_{13}$ plane. The damping coefficient of GFM-RES remains constant, which differs from the case of GFL-RES derived in the subsequent section.

Note that the expression of $P_{\mathrm{FM}}$ in (21) is given by (18) and (19), with their state variables determined via (12)-(16). Consequently, the effects of the *Q-V* control loop of GFM-RES and the HVRT/LVRT control loop of GFL-RES are inherently included. The subsequent derivations follow the same principle.

*2) GFL-RES*: According to the control principle of the PLL shown in Fig. 2, the dynamic equation of GFL-RES can be expressed as:

$$\ddot{\delta}_{13}=\underbrace{K_{\mathrm{i,PLL}}u_{\mathrm{FL,q}}}_{-F_{\mathrm{FL,p}}}+\underbrace{K_{\mathrm{p,PLL}}\dot{u}_{\mathrm{FL,q}}}_{F_{\mathrm{FL,d}}} \tag{23}$$

where $F_{\mathrm{FL,p}}$ and $F_{\mathrm{FL,d}}$ indicates the potential force and damping force of GFL-RES, respectively.

When GFM-RES operates under normal control mode, $u_{\mathrm{FL,q}}$ is calculated as:

$$\underbrace{u_{\mathrm{FL,q}}}_{n=1\sim3}=\begin{bmatrix} M_{3,1}U_{\mathrm{SYS}}\sin\left(\varphi_{M_{3,1}}-\delta_{13}\right)+\\ M_{3,2}U_{\mathrm{FM}}\sin\left(\varphi_{M_{3,2}}+\delta_{12}-\delta_{13}\right)+\\ M_{3,3}I_{\mathrm{FL}}\sin\left(\varphi_{M_{3,3}}+\varphi_{\mathrm{FL}}\right)\end{bmatrix} \tag{24}$$

Differentiate (24) and simplify the result, it is found that $\dot{u}_{\mathrm{FL,q}}$ is the inner product of the damping vector and the angle vector, indicating that the damping force of GFL-RES is a two-dimensional form.

$$\dot{u}_{\mathrm{FL,q}}=\frac{-1}{K_{\mathrm{p,PLL}}}\begin{bmatrix}D_{\mathrm{FL,12}}\\ D_{\mathrm{FL,13}}\end{bmatrix}\cdot\begin{bmatrix}\dot{\delta}_{12}\\ \dot{\delta}_{13}\end{bmatrix}=\frac{1}{K_{\mathrm{p,PLL}}}\begin{bmatrix}F_{\mathrm{FL,d}}^{12}\\ F_{\mathrm{FL,d}}^{13}\end{bmatrix} \tag{25}$$

$$\underbrace{\frac{D_{\mathrm{FL,12}}}{K_{\mathrm{p,PLL}}}}_{n=1\sim3}=\begin{bmatrix}-M_{3,2}\dfrac{\partial U_{\mathrm{FM}}}{\partial\delta_{12}}\sin\left(\varphi_{M_{3,2}}+\delta_{12}-\delta_{13}\right)-\\ M_{3,2}U_{\mathrm{FM}}\cos\left(\varphi_{M_{3,2}}+\delta_{12}-\delta_{13}\right)-\\ M_{3,3}\dfrac{\partial I_{\mathrm{FL}}}{\partial\delta_{12}}\sin\left(\varphi_{M_{3,3}}+\varphi_{\mathrm{FL}}\right)-\\ M_{3,3}\dfrac{\partial\varphi_{\mathrm{FL}}}{\partial\delta_{12}}I_{\mathrm{FL}}\cos\left(\varphi_{M_{3,3}}+\varphi_{\mathrm{FL}}\right)\end{bmatrix} \tag{26}$$

$$\underbrace{\frac{D_{\mathrm{FL,13}}}{K_{\mathrm{p,PLL}}}}_{n=1\sim3}=\begin{bmatrix}M_{3,1}U_{\mathrm{SYS}}\cos\left(\varphi_{M_{3,1}}-\delta_{13}\right)-\\ M_{3,2}\dfrac{\partial U_{\mathrm{FM}}}{\partial\delta_{13}}\sin\left(\varphi_{M_{3,2}}+\delta_{12}-\delta_{13}\right)+\\ M_{3,2}U_{\mathrm{FM}}\cos\left(\varphi_{M_{3,2}}+\delta_{12}-\delta_{13}\right)-\\ M_{3,3}\dfrac{\partial I_{\mathrm{FL}}}{\partial\delta_{13}}\sin\left(\varphi_{M_{3,3}}+\varphi_{\mathrm{FL}}\right)-\\ M_{3,3}\dfrac{\partial\varphi_{\mathrm{FL}}}{\partial\delta_{13}}I_{\mathrm{FL}}\cos\left(\varphi_{M_{3,3}}+\varphi_{\mathrm{FL}}\right)\end{bmatrix} \tag{27}$$

Once GFM-RES operates under current saturation mode, the expression for $u_{\mathrm{FL,q}}$ is given by (28), and the damping term satisfies (29) and (30).

$$\underbrace{u_{\mathrm{FL,q}}}_{n=4\sim6}=\begin{bmatrix}M_{3,1}U_{\mathrm{SYS}}\sin\left(\varphi_{M_{3,1}}-\delta_{13}\right)+\\ M_{3,2}I_{\mathrm{FM}}^{\mathrm{sa}}\sin\left(\varphi_{M_{3,2}}+\varphi_{\mathrm{FM}}^{\mathrm{sa}}+\delta_{12}-\delta_{13}\right)+\\ M_{3,3}I_{\mathrm{FL}}\sin\left(\varphi_{M_{3,3}}+\varphi_{\mathrm{FL}}\right)\end{bmatrix} \tag{28}$$

$$\underbrace{\frac{D_{\mathrm{FL,12}}}{K_{\mathrm{p,PLL}}}}_{n=4\sim6}=\begin{bmatrix}-M_{3,2}I_{\mathrm{FM}}^{\mathrm{sa}}\cos\left(\varphi_{M_{3,2}}+\varphi_{\mathrm{FM}}^{\mathrm{sa}}+\delta_{12}-\delta_{13}\right)\\ -M_{3,3}\dfrac{\partial I_{\mathrm{FL}}}{\partial\delta_{12}}\sin\left(\varphi_{M_{3,3}}+\varphi_{\mathrm{FL}}\right)-\\ M_{3,3}\dfrac{\partial\varphi_{\mathrm{FL}}}{\partial\delta_{12}}I_{\mathrm{FL}}\cos\left(\varphi_{M_{3,3}}+\varphi_{\mathrm{FL}}\right)\end{bmatrix} \tag{29}$$

$$\underbrace{\frac{D_{\mathrm{FL,13}}}{K_{\mathrm{p,PLL}}}}_{n=4\sim6}=\begin{bmatrix}M_{3,1}U_{\mathrm{SYS}}\cos\left(\varphi_{M_{3,1}}-\delta_{13}\right)+\\ M_{3,2}I_{\mathrm{FM}}^{\mathrm{sa}}\cos\left(\varphi_{M_{3,2}}+\varphi_{\mathrm{FM}}^{\mathrm{sa}}+\delta_{12}-\delta_{13}\right)\\ -M_{3,3}\dfrac{\partial I_{\mathrm{FL}}}{\partial\delta_{13}}\sin\left(\varphi_{M_{3,3}}+\varphi_{\mathrm{FL}}\right)-\\ M_{3,3}\dfrac{\partial\varphi_{\mathrm{FL}}}{\partial\delta_{13}}I_{\mathrm{FL}}\cos\left(\varphi_{M_{3,3}}+\varphi_{\mathrm{FL}}\right)\end{bmatrix} \tag{30}$$

Similar to the GFM-RES, the energy transformation and dissipation of GFL-RES satisfies $E_{\mathrm{FL,k}}=E_{\mathrm{FL,p}}+E_{\mathrm{FL,d}}$, where:

$$E_{\mathrm{FM,k}}=\frac{1}{2}\left(\dot{\delta}_{13}\right)^{2} \tag{31}$$

$$E_{\mathrm{FL,p}}=\int_{C}(-K_{\mathrm{i,PLL}}u_{\mathrm{FL,q}})\mathrm{d}\zeta \tag{32}$$

$$E_{\mathrm{FL,d}}=\int_{C}-\left(D_{\mathrm{FL,12}}\dot{\delta}_{12}+D_{\mathrm{FL,13}}\dot{\delta}_{13}\right)\mathrm{d}\zeta \tag{33}$$

### *C. Dominant Instability Mode and Dynamic Interaction in the GFM/GFL-RES Hybrid Systems*

Based on the distribution of $R_n\,(n=1\sim6)$, potential force, and damping force solved in Section III.B, the instability mode of the hybrid system and interaction mechanism between GFM/GFL-RES are analyzed in this section.

*1) Dominant Instability Mode*: Considering the coupling between state variables and control combination region $R_n$, the characteristics of the potential forces $F_{\mathrm{FM,p}}$ and $F_{\mathrm{FL,p}}$ with respect to $\delta_{12}$ and $\delta_{13}$ are visualized in Fig. 6 and Fig. 7, respectively. The subfigure (b) can be interpreted as a top view of subfigure (a), in which the regions filled with white solid stripes (i.e., $R_{\mathrm{FM,p}}^{+}$ and $R_{\mathrm{FL,p}}^{+}$) indicate that the potential forces provide a braking effect in these areas. Additionally, the black solid curves in subfigure (b) denote the zero-level contour of the potential force, which can be realized as the extension of SEP and UEP of a two-machine system to a multi-machine system.

For a stable three-machine system, it is essential to ensure that both GFM-RES and GFL-RES asymptotically return to the initial SEP after a disturbance. This requirement implies that the initial SEP must satisfy both $\partial P_{\mathrm{FM}}/\partial\delta_{12}>0$ and $\partial P_{\mathrm{FL}}/\partial\delta_{13}>0$. Hence, the SEP in the three-machine system can be expressed as:

$$\mathrm{SEP}\in\left\{\vartheta\in\Theta\left|\begin{array}{l}\dfrac{\partial P_{\mathrm{FM}}(\vartheta)}{\partial\delta_{12}}>0,\ \dfrac{\partial u_{\mathrm{FL,q}}(\vartheta)}{\partial\delta_{13}}<0,\\ P_{\mathrm{FM}}(\vartheta)=P_{\mathrm{FM}}^{\mathrm{ref}},\ u_{\mathrm{FL,q}}(\vartheta)=0\end{array}\right.\right\} \tag{34}$$

where $\vartheta$ is defined as a point in the $\delta_{12}$-$\delta_{13}$ phase space. $\Theta\in\mathbb{R}^2$ indicates the operation region of the system. Furthermore, the UEP in the GFM/GFL-RES hybrid system can be expressed as:

$$\mathrm{UEP}\in\left\{\vartheta\in\Theta\left|\begin{array}{l}\left[\dfrac{\partial P_{\mathrm{FM}}(\vartheta)}{\partial\delta_{12}}<0\vee\dfrac{\partial u_{\mathrm{FL,q}}(\vartheta)}{\partial\delta_{13}}>0\right],\\ P_{\mathrm{FM}}(\vartheta)=P_{\mathrm{FM}}^{\mathrm{ref}},\ u_{\mathrm{FL,q}}(\vartheta)=0\end{array}\right.\right\} \tag{35}$$

As shown in Fig. 6(b), there are two SEP sets (i.e., $\Gamma_{\mathrm{FM}}^{\mathrm{SEP,1}}$ and $\Gamma_{\mathrm{FM}}^{\mathrm{SEP,2}}$) and one UEP set $\Gamma_{\mathrm{FM}}^{\mathrm{UEP}}$. $\Gamma_{\mathrm{FM}}^{\mathrm{SEP,1}}$ and $\Gamma_{\mathrm{FM}}^{\mathrm{SEP,2}}$ correspond to the normal control mode and current saturation mode of GFM-RES, respectively. In contrast, $F_{\mathrm{FL,p}}$ contains only one SEP set $\Gamma_{\mathrm{FL}}^{\mathrm{SEP}}$ and one UEP set $\Gamma_{\mathrm{FL}}^{\mathrm{UEP}}$, as shown in Fig. 7(b).

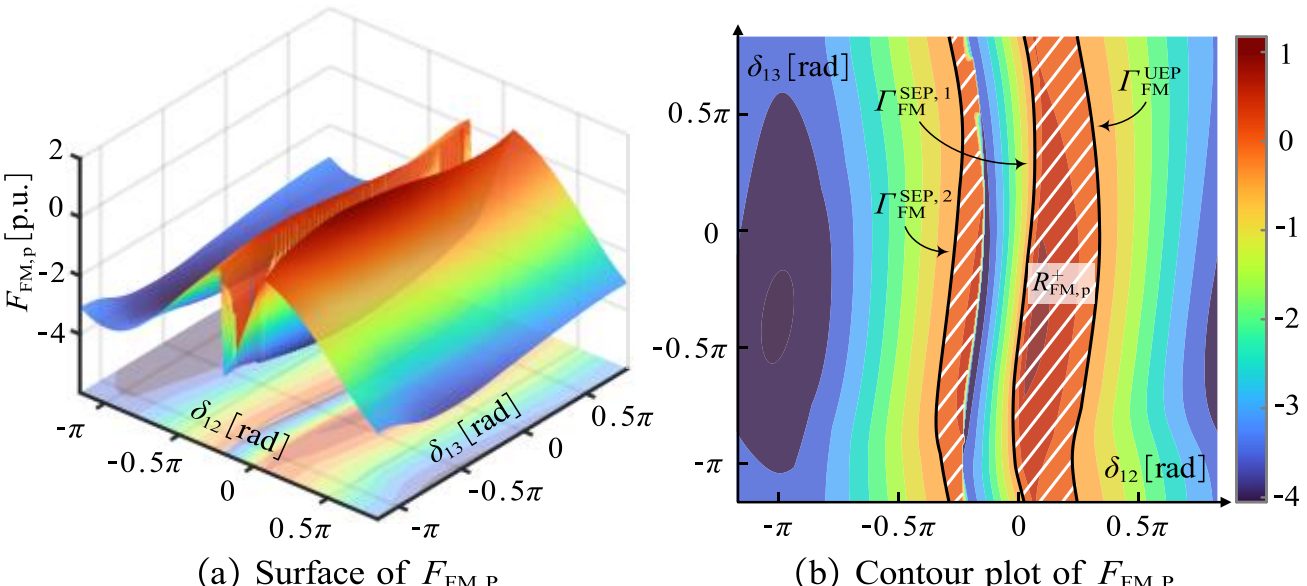

(a) Surface of $F_{\mathrm{FM,P}}$ (b) Contour plot of $F_{\mathrm{FM,P}}$

Fig. 6. Characteristics of the GFM-RES potential forces $F_{\mathrm{FM,p}}$.

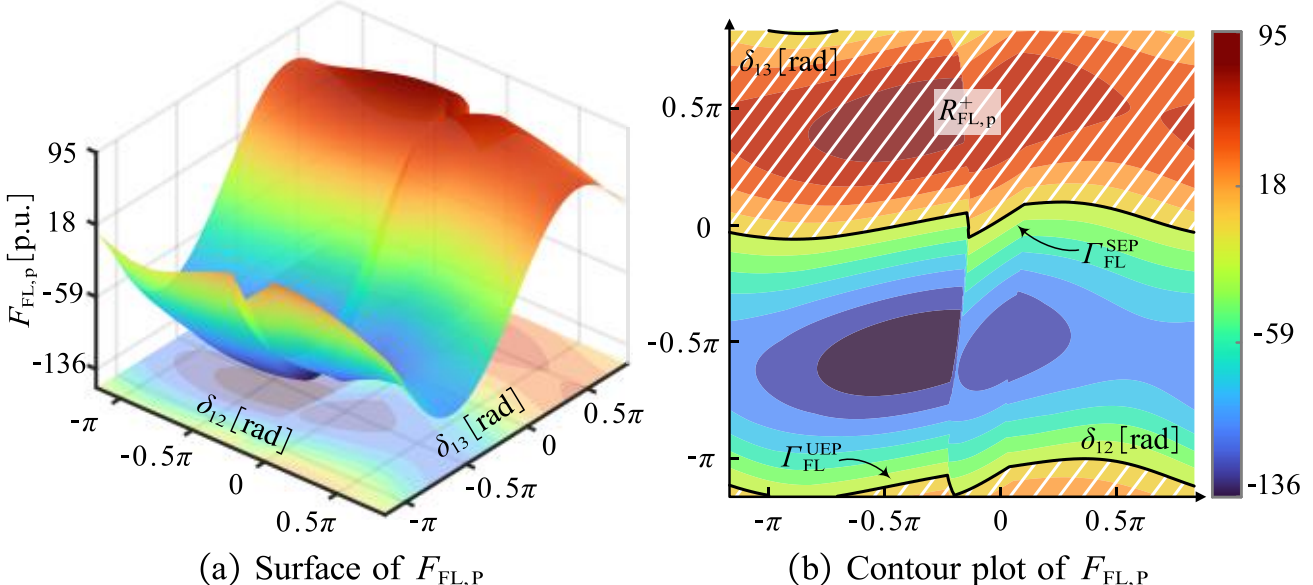

(a) Surface of $F_{\mathrm{FL,P}}$ (b) Contour plot of $F_{\mathrm{FL,P}}$

Fig. 7. Characteristics of the GFL-RES potential forces $F_{\mathrm{FL,p}}$.

Fig. 8 presents the first-swing instability modes of the GFM/GFL-RES hybrid system with the critical angular velocity equal to $\omega_0$. The intersections of the curve $\Gamma_{\mathrm{FL}}^{\mathrm{SEP}}$ with curves $\Gamma_{\mathrm{FM}}^{\mathrm{SEP,1}}$ and $\Gamma_{\mathrm{FM}}^{\mathrm{SEP,2}}$ corresponding to $\mathrm{SEP}_1$ and $\mathrm{SEP}_2$, respectively, where $\mathrm{SEP}_1$ is the initial SEP and $\mathrm{SEP}_2$ is an abnormal SEP induced by the current saturation mode of GFM-RES. In the steady state, the system stabilizes at $\mathrm{SEP}_1$. If the operating point first reaches $\Gamma_{\mathrm{FM}}^{\mathrm{UEP}}$ during the transient process, the operating point will continue to accelerate in the direction of increasing $\delta_{12}$, which subsequently leads to the GFM-RES dominated instability, as illustrated by trajectories 1 and 2 in Fig. 8. In contrast, when the operating point first contacts with $\Gamma_{\mathrm{FL}}^{\mathrm{UEP}}$, then the first-swing instability dominated by GFL-RES will occur, as shown by trajectories 3 and 4.

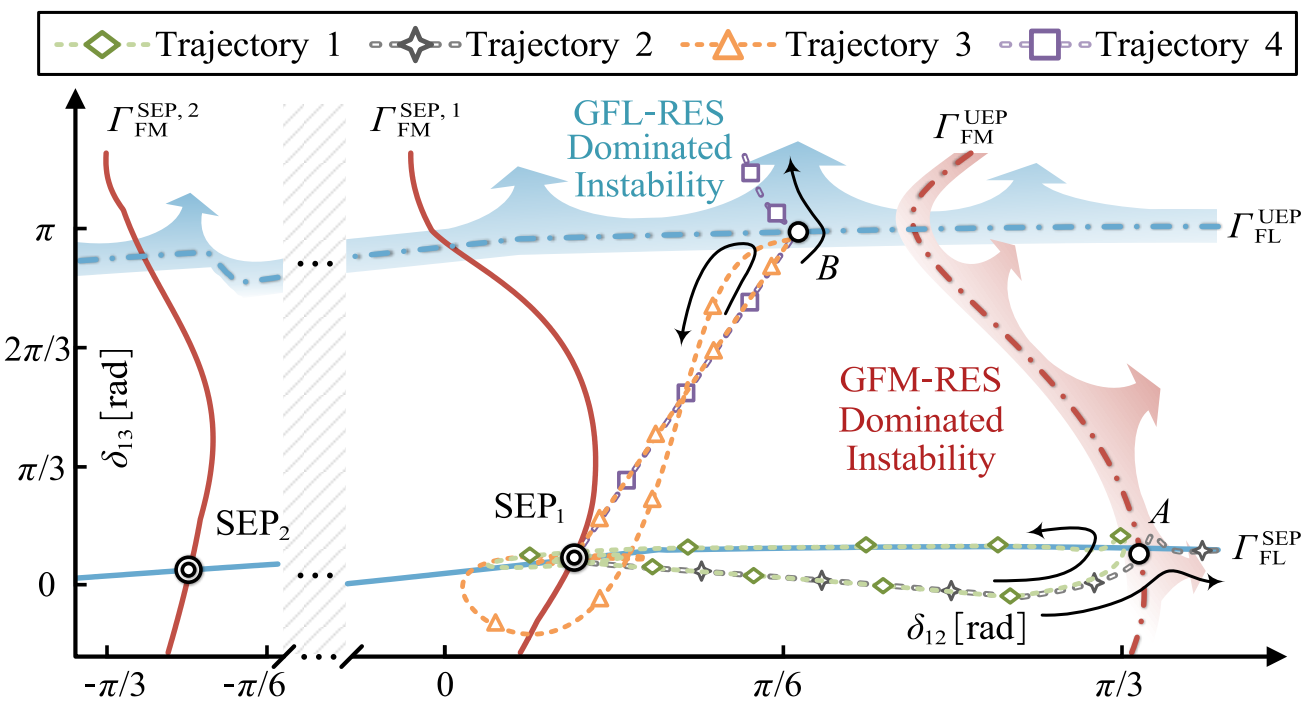

Fig. 8. Dominant instability mode of the GFM/GFL-RES hybrid system.

Based on the above analysis, the dominant instability principle in the first-swing can be given as follows:

$$f_{\vartheta}=\begin{cases}\text{GFM-RES}, & \vartheta\in\mathcal{N}(\Gamma_{\mathrm{FM}}^{\mathrm{UEP}})\cap[\Theta\setminus R_{\mathrm{FM,p}}^{+}]\\ \text{GFL-RES}, & \vartheta\in\mathcal{N}(\Gamma_{\mathrm{FL}}^{\mathrm{UEP}})\cap[\Theta\setminus R_{\mathrm{FL,p}}^{+}]\end{cases} \tag{36}$$

where $f_{\vartheta}$ indicates the dominant instability flag. $\mathcal{N}(\cdot)$ is defined as the neighborhoods function. When $\vartheta$ reaches the region $\mathcal{N}(\Gamma_{\mathrm{FM}}^{\mathrm{UEP}})\cap[\Theta\setminus R_{\mathrm{FM,p}}^{+}]$ with critical angular velocity, the GFM-RES is identified as the dominant source of instability. A similar conclusion can be drawn in the opposite case.

*2) Dynamic Interaction*: As illustrated in Section III.B, the potential forces $F_{\mathrm{FM,p}}$ and $F_{\mathrm{FL,p}}$ are coupled through external characteristics of sources (i.e., the power flow behavior of the network). However, the damping coupling mechanism between GFM-RES and GFL-RES is relatively more complex. According to (17), the damping coefficient $D_{\mathrm{FM}}$ remains constant throughout the operation, implying that the damping term of GFM-RES exhibits a linear relationship with $\dot{\delta}_{12}$, and the GFL-RES affects the GFM-RES only via the potential force. Nevertheless, as illustrated in (25)~(27), (29), and (30), the damping characteristics of GFL-RES are inherently two-dimensional, with distinct damping components associated with $\dot{\delta}_{12}$ and $\dot{\delta}_{13}$. Moreover, the corresponding damping coefficients vary as the implicit functions of $\delta_{12}$ and $\delta_{13}$, respectively.

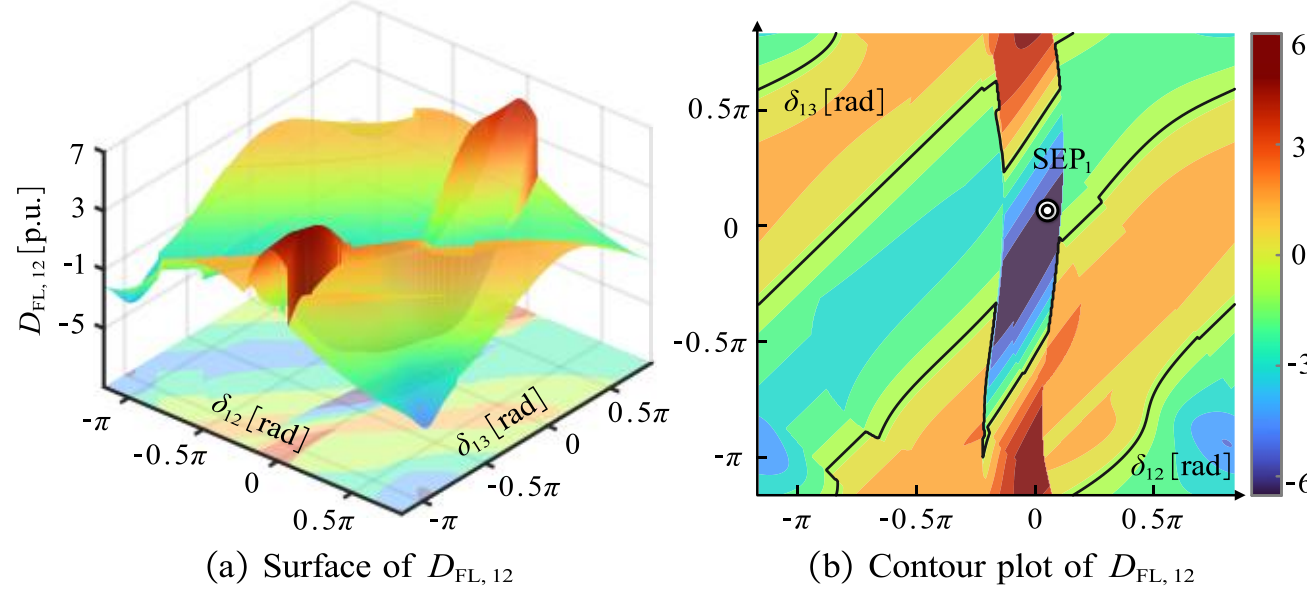

(a) Surface of $D_{\mathrm{FL,12}}$ (b) Contour plot of $D_{\mathrm{FL,12}}$

Fig. 9. Characteristics of the damping coefficient $D_{\mathrm{FL,12}}$.

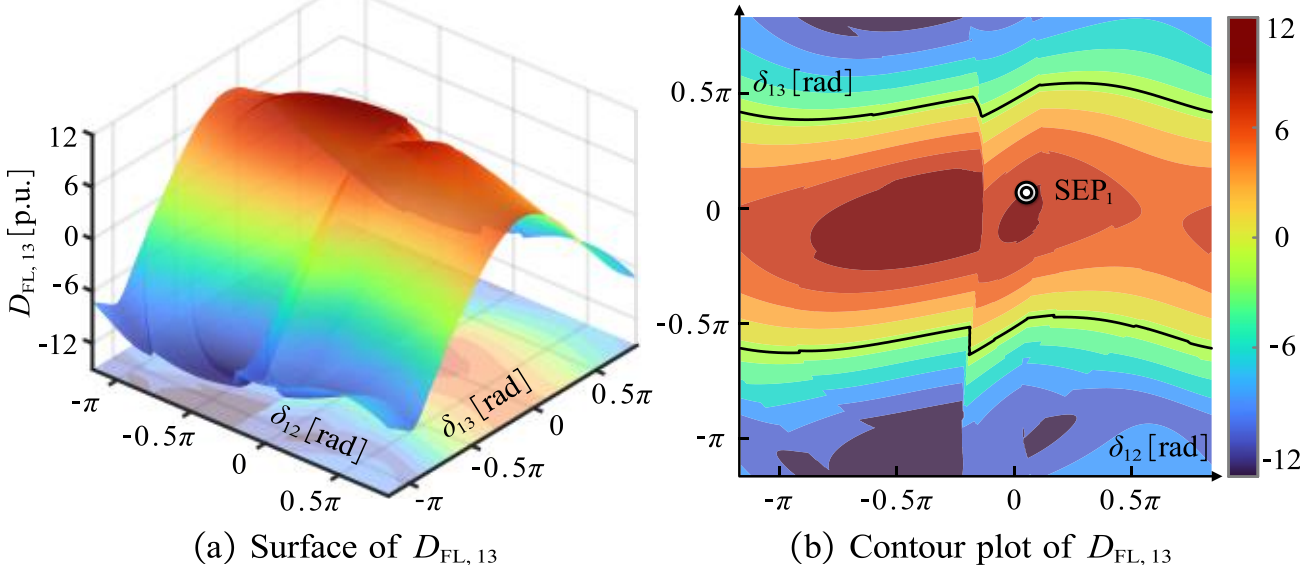

(a) Surface of $D_{\mathrm{FL,13}}$ (b) Contour plot of $D_{\mathrm{FL,13}}$

Fig. 10. Characteristics of the damping coefficient $D_{\mathrm{FL,13}}$.

Fig. 9 and Fig. 10 present the global distribution of damping coefficients $D_{\mathrm{FL,12}}$ and $D_{\mathrm{FL,13}}$ under six control combinations. The black solid curve indicates the zero-level contour of the damping coefficient. It is observed that the $D_{\mathrm{FL,12}}$ and $D_{\mathrm{FL,13}}$ exhibit distinct distribution patterns in the $\delta_{12}$-$\delta_{13}$ space. Specifically, $D_{\mathrm{FL,12}}$ tends to align along the direction of $\delta_{12}=\delta_{13}$, indicating that both $\delta_{12}$ and $\delta_{13}$ have significant influence on $D_{\mathrm{FL,12}}$. Additionally, $D_{\mathrm{FL,13}}$ is primarily distributed along the $\delta_{12}$ axis, suggesting that the impact of $\delta_{12}$ is limited, while $\delta_{13}$ plays a dominant role in determining $D_{\mathrm{FL,13}}$.

As shown in Fig. 9(b), it can be seen that $D_{\mathrm{FL,12}}$ remains negative over a relatively wide range surrounding SEP$_1$, which implies that this damping term will exert a destabilizing effect when the corresponding dynamic equations contain high-order derivatives of $\delta_{12}$. In contrast, as seen from (23) and (25), the damping term $F_{\mathrm{FL,d}}^{12}$ (i.e., $-D_{\mathrm{FL,12}}\dot{\delta}_{12}$) ultimately exerts impact on $\ddot{\delta}_{13}$, rather than $\ddot{\delta}_{12}$. Hence, both spatial and temporal dependence are imparted to $F_{\mathrm{FL,d}}^{12}$ through this cross-coupled damping mechanism, as it evolves with both system state and time. In other words, this observation reveals that the stabilizing influence exerted by GFM-RES on GFL-RES possesses a spatiotemporal distribution characteristic.

Based on the above analysis, the criterion for evaluating the impact of the GFM-RES on the transient stability of GFL-RES can be expressed as follows:

$$f_{D_{12}}=\begin{cases}1\ , & \vartheta\in R_{\mathrm{FL,d}}^{12,+}\\ -1\ , & \vartheta\in R_{\mathrm{FL,d}}^{12,-}\end{cases} \tag{37}$$

where

$$R_{\mathrm{FL,d}}^{12,+}=\{\vartheta\in\Theta\,|\,\mathrm{sgn}(v)D_{\mathrm{FL,12}}(\vartheta)>0\} \tag{38}$$

$$R_{\mathrm{FL,d}}^{12,-}=\{\vartheta\in\Theta\,|\,\mathrm{sgn}(v)D_{\mathrm{FL,12}}(\vartheta)<0\} \tag{39}$$

$$\mathrm{sgn}(v)=\begin{cases}1\ , & v>0\\ -1\ , & v<0\end{cases}\ ,\ v=\dot{\delta}_{12}\dot{\delta}_{13} \tag{40}$$

In (37)through (40), $f_{D_{12}}$ indicate the flag that identifies whether the $F_{\mathrm{FL,d}}^{12}$ is positive or negative. $R_{\mathrm{FL,d}}^{12,+}$ and $R_{\mathrm{FL,d}}^{12,-}$ denote the regions where $F_{\mathrm{FL,d}}^{12}$ is positive or negative, respectively. $\mathrm{sgn}(\cdot)$ denotes a sign function, which is used to determine the sign relationship between $\dot{\delta}_{12}$ and $\dot{\delta}_{13}$. When $f_{D_{12}}=1$, $F_{\mathrm{FL,d}}^{12}$ exhibits positive damping force characteristic, thus enhancing the transient stability of GFL-RES. Conversely, when $f_{D_{12}}=-1$, the negative damping behavior of $F_{\mathrm{FL,d}}^{12}$ tends to accelerate the GFL-RES, thereby the transient performance of GFL-RES is deteriorated.

Based on the proposed systematic analysis framework, the majority of the transient instability mechanisms in the GFM/GFL-RES hybrid system can be uncovered.

## IV. Simulation Validations

The electromagnetic transient simulation software PSCAD/EMTDC is employed to verify the effectiveness of the multi-machine stability analysis framework developed in section III under a scenario involving significant switching dynamics.

The studied system is shown in Fig. 3, in which a three-phase to ground short-circuit fault with a grounding resistance of 1 Ω is applied. The fault duration is 1.2 s and the line impedances are given as $Z_{14}=(2+j9.3)\times10^{-2}$, $Z_{24}=(0.7+j5.5)\times10^{-2}$, and $Z_{34}=(1+j6.5)\times10^{-2}$, in per unit. For simplicity of analysis, the load is modeled solely as a constant impedance. In the time-domain simulation system, both $\delta_{12}$ and $\delta_{13}$ are calculated with reference to the main grid phase, which can be obtained from the phase measurement unit at the PCC. The remaining parameters are listed in TABLE II. Based on the transient instability pattern illustrated in Fig. 11, the correctness of the derived control combination region for the GFM/GFL-RES hybrid system and the proposed dynamic interaction mechanism between GFM/GFL-RES will be validated.

TABLE II

Parameters of GFM/GFL-RES and the Grid

| Parameter | Symbol | Value |
|---|---|---|
| Base active power | $S_{\mathrm{base}}$ | 100 MW |
| Base voltage | $V_{\mathrm{base}}$ | 230 kV |
| Base frequency | $f_{\mathrm{base}}$ | 50 Hz |
| System voltage | $U_{\mathrm{SYS}}$ | 1.0 p.u. |
| Load power | $[P_{\mathrm{load}}, Q_{\mathrm{load}}]$ | [2.49, 0.62] p.u. |
| GFM-RES power reference | $[P_{\mathrm{FM}}^{\mathrm{ref}}, Q_{\mathrm{FM}}^{\mathrm{ref}}]$ | [1.68, 0.21] p.u. |
| GFM-RES setting voltage | $u_{\mathrm{FM,0}}$ | 1.01 p.u. |
| GFM-RES virtual inertia | $J_{\mathrm{FM}}$ | 0.5 |
| GFM-RES virtual damping | $D_{\mathrm{FM}}$ | 1 |
| GFM-RES $Q$-$V$ droop | $K_Q$ | 0.5 |

| Parameter | Symbol | Value |
|---|---|---|
| GFM-RES current saturation phase | $\varphi_{\mathrm{FM}}^{\mathrm{sa}}$ | -$\pi$/18 rad |
| GFM-RES max current ratio | $I_{\mathrm{FM}}^{\max}/I_{\mathrm{FM},0}$ | 1.5 |
| GFL-RES power reference | $[P_{\mathrm{FL}}^{\mathrm{ref}}, Q_{\mathrm{FL}}^{\mathrm{ref}}]$ | [1.39, 0.27] p.u. |
| GFL-RES PLL parameters | $[K_{\mathrm{p,PLL}}, K_{\mathrm{i,PLL}}]$ | [10, 100] |
| GFL-RES LVRT/HVRT parameters | $[K_{I}^{\mathrm{RT}}, K_{\varphi}^{\mathrm{RT}}]$ | [0.50, 2.46] |
| GFL-RES max current ratio | $I_{\mathrm{FL}}^{\max}/I_{\mathrm{FL},0}$ | 1.2 |

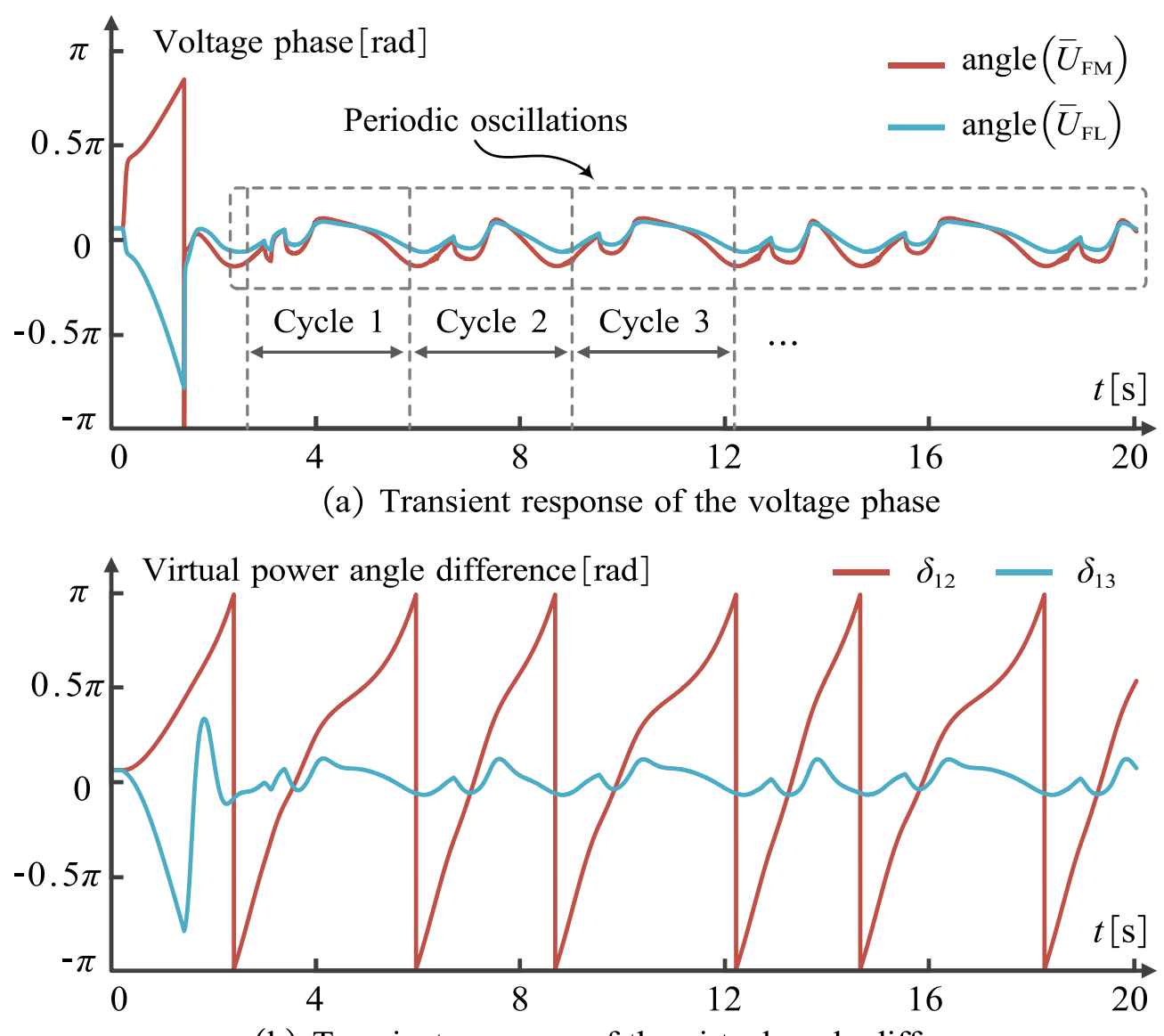

Fig. 11. Transient response of the GFM/GFL-RES hybrid system.

Fig. 11 presents the transient response of the system following the fault, in which the GFM-RES loses synchronization at the control level (see Fig. 11(b)) and exhibits periodic oscillations at the electrical level (see Fig. 11(a)). Taking the trajectory of operating point from $t = 0$ s to the end of Cycle 1, the corresponding theoretical control combinations are solved through (15), (16), and TABLE I, as illustrated in Fig. 12(a), while the control switching flags of GFM/GFL-RES is given in Fig. 12(b), (c), respectively. During the fault, the current saturation mode of GFM-RES is triggered to prevent overcurrent issues. Meanwhile, GFL-RES switches to the LVRT mode due to the voltage sag of the system. At this stage, both GFM/GFL-RESs are subjected to unbalanced force, driving them away from the initial SEP, corresponding to the transition from $\mathrm{SEP}_1$ to $A_1$ in Fig. 12(a). After fault clearance, the operating point moves from $A_1$ to $D_1$, during which the GFL-RES switches to the normal control mode at $B_1$ and switches back to the LVRT mode at $C_1$, each aligning with the theoretical boundary. The trajectory from $D_1$ to $G_1$ corresponding to Cycle 1 in Fig. 11(a), and the majority of the control switching events are consistent with the theoretical results, except for the transition form $E_{1,1}$ to $E_{1,2}$, which can be attributed to some uncontrollable transient characteristics.

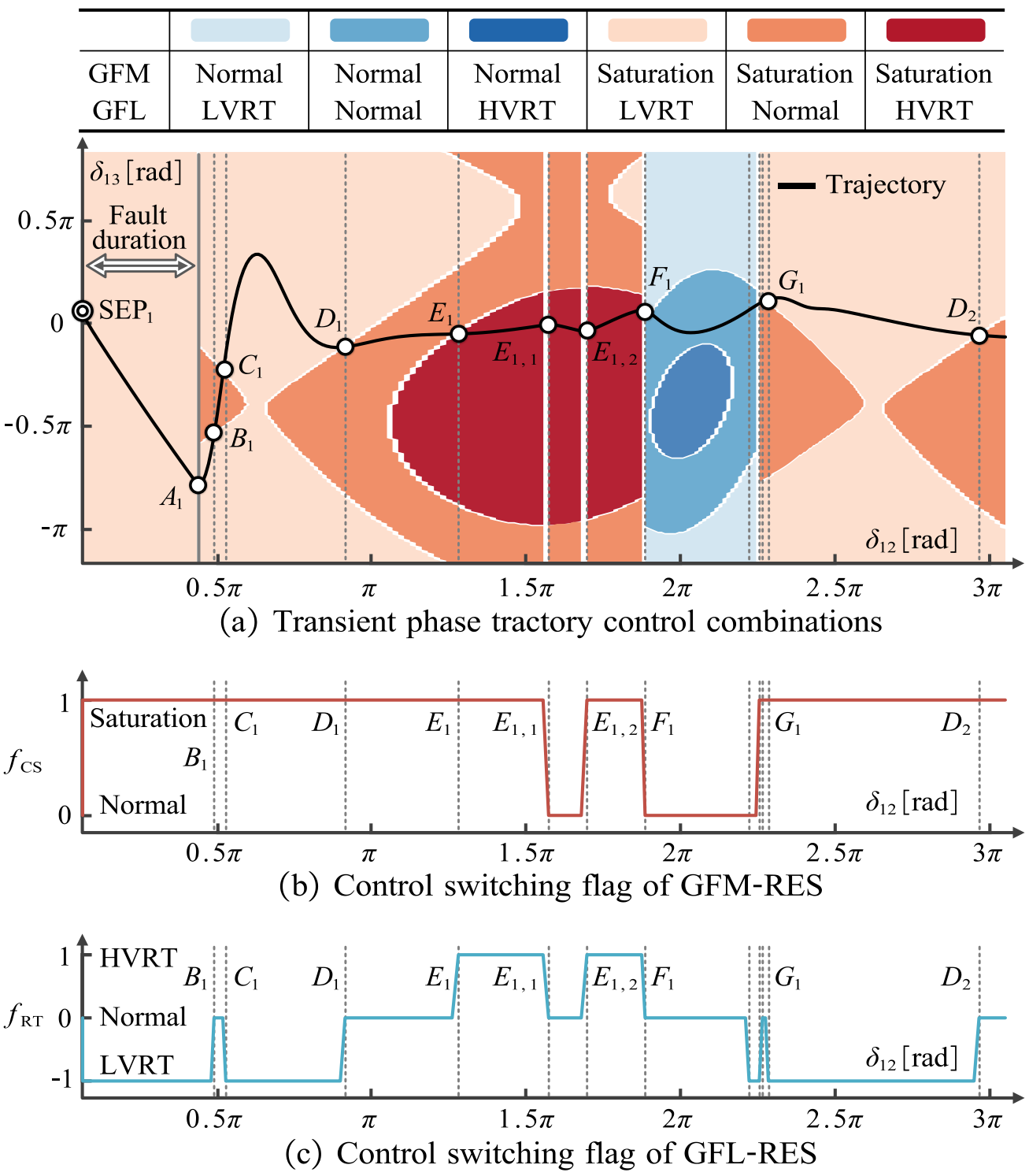

Fig. 12. Validation of derived control combination region in an instability case with significant switching dynamics.

Fig. 13 further validates the correctness of the derived control combination region and reveals the mechanism of periodic oscillations, as the theoretical voltage phase of GFM-RES matches the simulation result very well. The theoretical value obtained from the phasor model can be regarded as the reference voltage phase at the PCC, which may undergo sudden changes during control switching. However, due to the inherent inertia of the actual voltage phase, it does not instantly jump to the reference value during the transient process, as illustrated by the gray dashed boxes in Fig. 13. It also confirms that although $\delta_{12}$ increasing monotonically at the control level, the voltage phase at the PCC is theoretically expected to exhibit periodic oscillations under the influence of control switching.

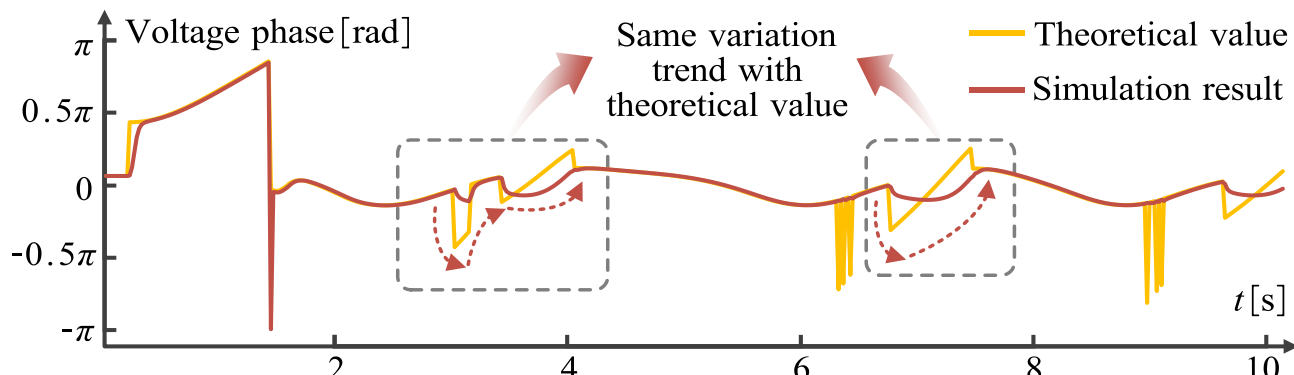

Fig. 13. Validation of voltage phase between theoretical and simulated results.

The dynamic characteristics of the GFL-RES damping force are presented in Fig. 14, where subfigure (a) further validates the correctness of the derived system dynamics. As observed from Fig. 14(b), the impact of the damping force component $F_{\mathrm{FL,d}}^{12}$ on GFL-RES exhibits continuous transitions over time. The underlying mechanism that causes the sign-varying characteristic of $F_{\mathrm{FL,d}}^{12}$ is shown in Fig. 14(c) through Fig. 14 (e), in which subfigure (c) reflects the joint effects of subfigure (d) and (e).

According to the criterion proposed in (37) through (40), the dynamic influence of GFM-RES on GFL-RES manifests in both temporal and spatial domains. In the temporal domain, it is affected by the sign consistency of angular velocities of GFM/GFL-RES, corresponding to the red dashed line labeled *A* to *C*, *E* to *H*, and *J* in Fig. 14(d). On the spatial level, the impact is determined by the nonlinear characteristics of $D_{\mathrm{FL},12}$ in the $\delta_{12}$-$\delta_{13}$ phase plane (see Fig. 9), as shown by the cyan dash-dotted lines *D* and *I* in Fig. 14(e).

Fig. 15 conceptually depicts the dynamic coupling of GFM-RES and GFL-RES from their perspectives, where self-correlations are represented by solid arrows and cross-correlations are indicated by dashed arrows. In this abstraction, the GFM-RES and GFL-RES are represented by a red sphere and a cyan sphere, respectively, each moving along its operating trajectory. Both GFM/GFL-RES's potential forces are associated with $\delta_{12}$ and $\delta_{13}$, as shown in (17)-(19), (23), (24), and (28). Accordingly, along any given trajectory, the potential energy of GFM/GFL-RES consists of self and mutual components, analogous to the springs illustrated in Fig. 15. Due to their heterogeneous synchronization mechanisms, the damping forces acting on GFM-RES and GFL-RES differ. The damping force on the GFM-RES depends solely on its own $\delta_{12}$ and is not directly related to the GFL-RES. In contrast, the damping force on the GFL-RES is influenced by $\delta_{12}$, $\delta_{13}$, $\dot{\delta}_{12}$, and $\dot{\delta}_{13}$, exhibiting both spatial and temporal variability.

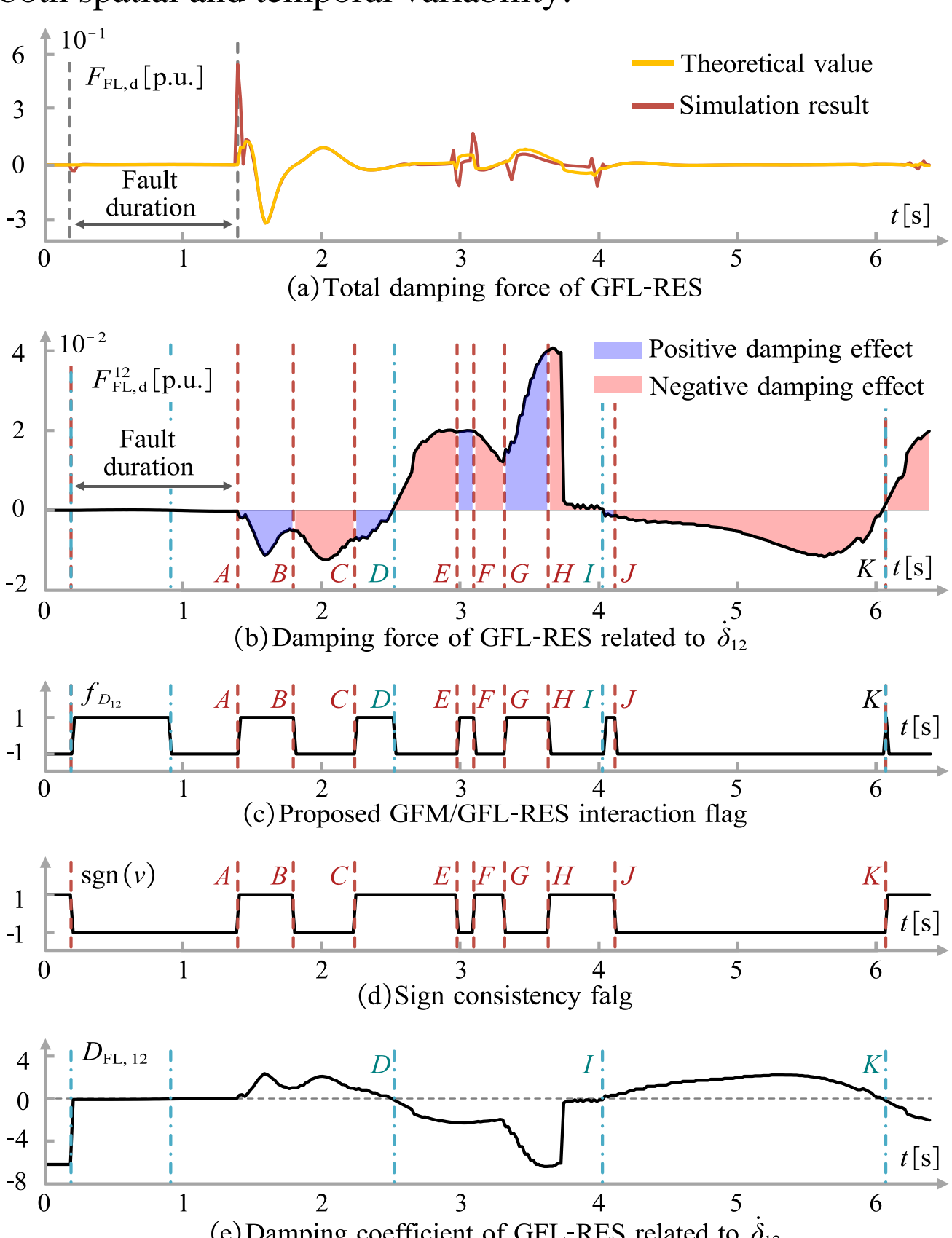

Fig. 14. Dynamic interaction during transient.

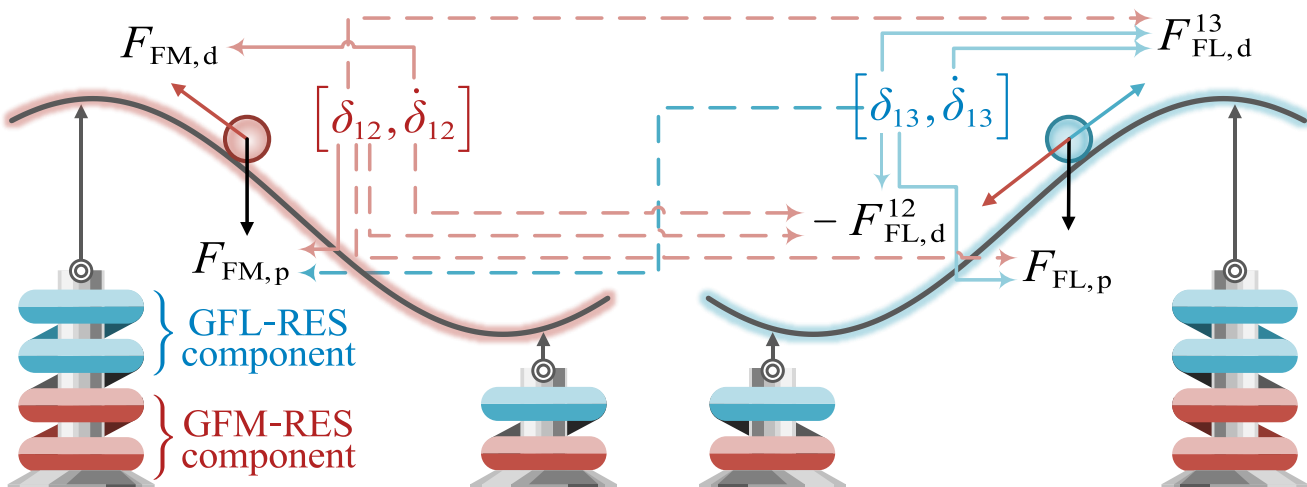

Fig. 15. Dynamic coupling between GFM/GFL-RES.

## V. Conclusion

This paper illustrates the dynamic interaction mechanisms and evolutions of the GFM/GFL-RES hybrid system considering multi-mode control switching. The six control combination regions are derived first to clarify the operating modes and switching boundaries of the system. Then, the dynamic interactions between GFM/GFL-RES are revealed from the perspectives of energy transformation and dissipation. Two instability modes in the first-swing dominated by GFM/GFL-RES are uncovered under specific conditions. Furthermore, it is identified that the damping force applied to GFL-RES inherently exhibits a two-dimensional characteristic, with the damping component provided by GFM-RES angular velocity demonstrating both spatial and temporal characteristics, further complicating the dynamic behavior of the GFM/GFL-RES hybrid system. The main conclusions are summarized as follows:

1) The control combinations of GFM/GFL-RES reshape the power flow distribution of the system, which intensifies the nonlinear coupling among state variables and may lead to discontinuities in power flow characteristics near switching boundaries.

2) The SEP of the hybrid system corresponds to the intersection of the SEP sets of GFM/GFL-RES, whereas the UEP sets are determined by the union of their respective UEP sets. Under critical conditions, if the operating point first crosses the unstable boundary of the GFM-RES, the instability is GFM-dominated. Otherwise, it is GFL-dominated.

3) The potential forces of GFM/GFL-RES are mutually coupled. Meanwhile, the damping force of GFL-RES behaves as a two-dimensional form, with one component governed by the angular velocity of GFM-RES exhibiting both temporal and spatial variability.

With respect to the operation of practical power systems, the contributions of this paper can serve as a reference in the following aspects:

1) During system planning and RES control design, the derivations presented herein can be utilized to enhance the first-swing stability of synchronized systems and increase stability margins.

2) The proposed dominant instability mode identification method can be applied to locate and recognize the dominant unstable device, clarify the underlying instability mechanism, and provide guidance for grid operators to perform precise converter tripping and rapid disturbance source isolation.

3) The interaction mechanisms investigated in this paper

provide a basis for enhancing the self-stabilizing capability of devices that are electrically distant through improved individual control of GFM/GFL-RESs. For devices that are electrically close, the results offer guidance for implementing coordinated control among GFM/GFL-RESs, thereby strengthening their mutual-stabilizing capability, mitigating adverse interactions, and ultimately improving the transient stability of next-generation power-electronics-dominated power systems.

Future work will focus on the stability analysis of bulk systems with diverse energy sources and the coordinated control among various RESs. In addition, Fig. 16 provides a conceptual illustration, based on the proposed framework, for analyzing the dynamics of multi-machine systems. We aim to highlight the extensibility of this study and outline potential challenges for future research:

1) *Diversity of energy sources*: In practical power systems, different types of energy sources may be equipped with heterogeneous control mechanisms, which in turn affect certain characteristics of the sources. For a typical GFL-RES, for instance, its damping behavior is directly influenced by voltage characteristics due to its *V-f* synchronization property. In a single-machine infinite-bus model, the spatial variation of damping is only reflected in the relative virtual power angle with respect to the system. However, when other sources are taken into account, the complexity of damping characteristics increases significantly, making the dynamic analysis of such devices more challenging. Nevertheless, the proposed modeling approach is generalizable, and by following the derivation procedure presented in this paper, the impacts of different synchronization loops and outer voltage/current loops [29] can be systematically analyzed.

2) *Load characteristics*: The load behavior in real systems is often complex, with some dynamic loads introducing rotational dynamics (e.g., induction motor loads), which further complicates overall system analysis. When the system frequency is close to the steady-state value, the phasor-based analytical model proposed in this paper can capture the characteristics of loads at both power flow and dynamic levels.

3) *Dimensionality of multi-machine systems*: As the number of devices increases, the dimensionality of the system's state space also expands. For example, in a system with $m$ sources, the operating region of relative power angles becomes ($m$-1)-dimensional. Furthermore, if each device is equipped with $k$ control strategies, the total number of control combination regions reaches $k \times m$. In this paper, control combination regions and stable/unstable equilibrium points are modeled as sets, and through high-dimensional mappings among these sets, the proposed method remains applicable to multi-machine systems to a certain extent.

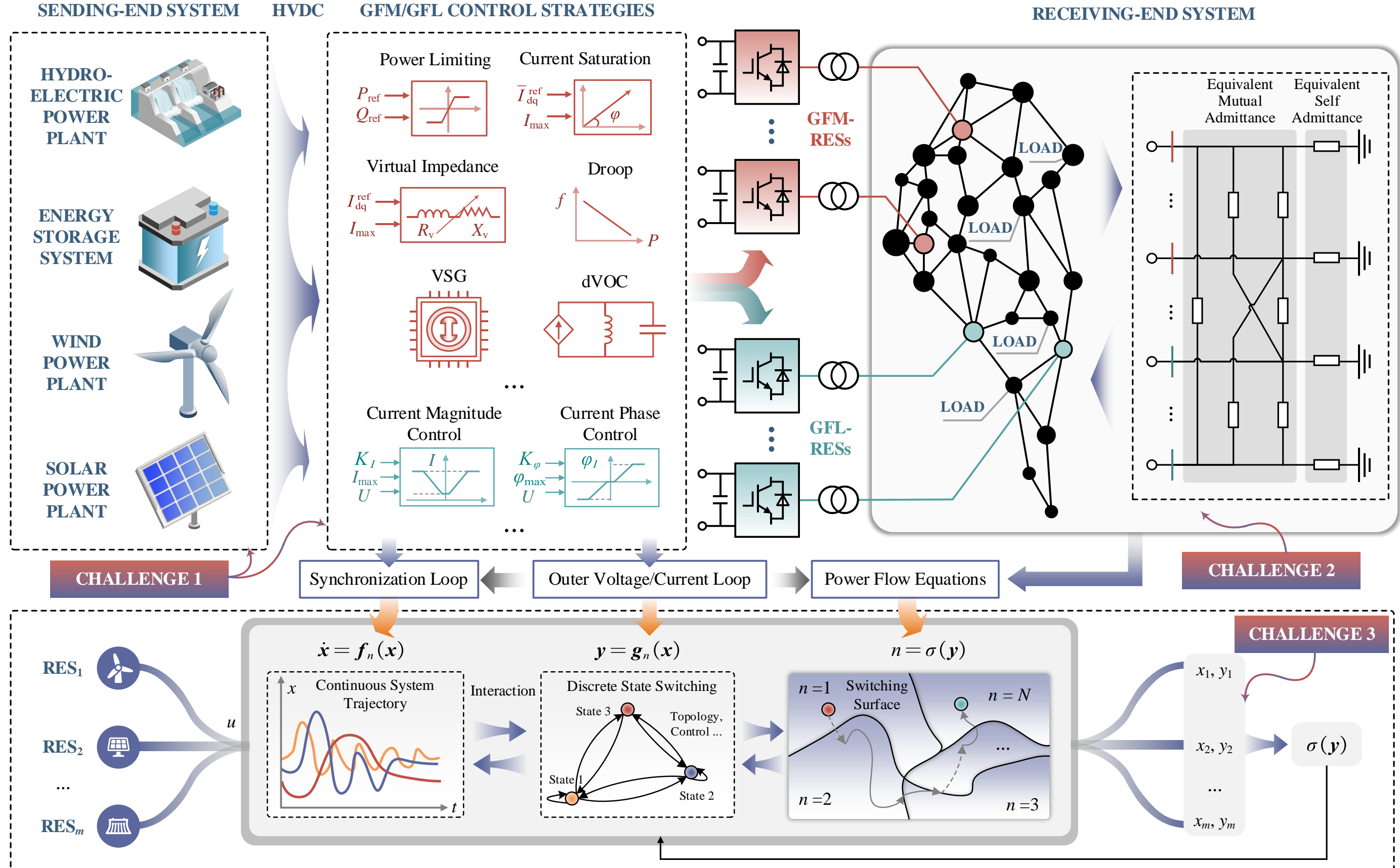

Fig. 16. Conceptual illustration of dynamic analysis in power systems with diverse energy sources. In analyzing the impact of different control strategies on system stability, the synchronization loops directly determine the system's dynamic mechanisms. In contrast, the outer voltage/current loops influence the output functions and power-flow characteristics, which in turn determine the system's control combination regions through the power-flow equations.

## APPENDIX

### A. Sensitivity Analysis of Transient Control Parameters in the Fault Recovery Stage

Fig. 17 illustrates the impact of different current saturation reference values of GFM-RES on the system's control combination regions. Subplots (a)-(c) correspond to current saturation phase angles of $\pi/4$, 0, and $-\pi/4$ rad, respectively, while subplots (d)-(f) present stacked comparisons of the control combination regions for $n$ = 4~6 from subplots (a)-(c). From subplots (a)-(f), it can be observed that variations in current phase angle primarily cause a shift of the control combination region along the $\delta_{12}$ axis. As the current reference phase decreases gradually from $\pi/4$ to $-\pi/4$, the control combination regions indexed by $n$ = 4~6 exhibit a left-to-right translation. Since current saturation causes the GFM-RES to shift from a voltage source to a current source, the switching conditions become inconsistent. Under such circumstances, an improper current reference phase may lead to continuous current saturation [28], as shown in Fig. 17(a), where GFM-RES abnormally operates in a reactive power absorption mode, deteriorating the system's transient performance.

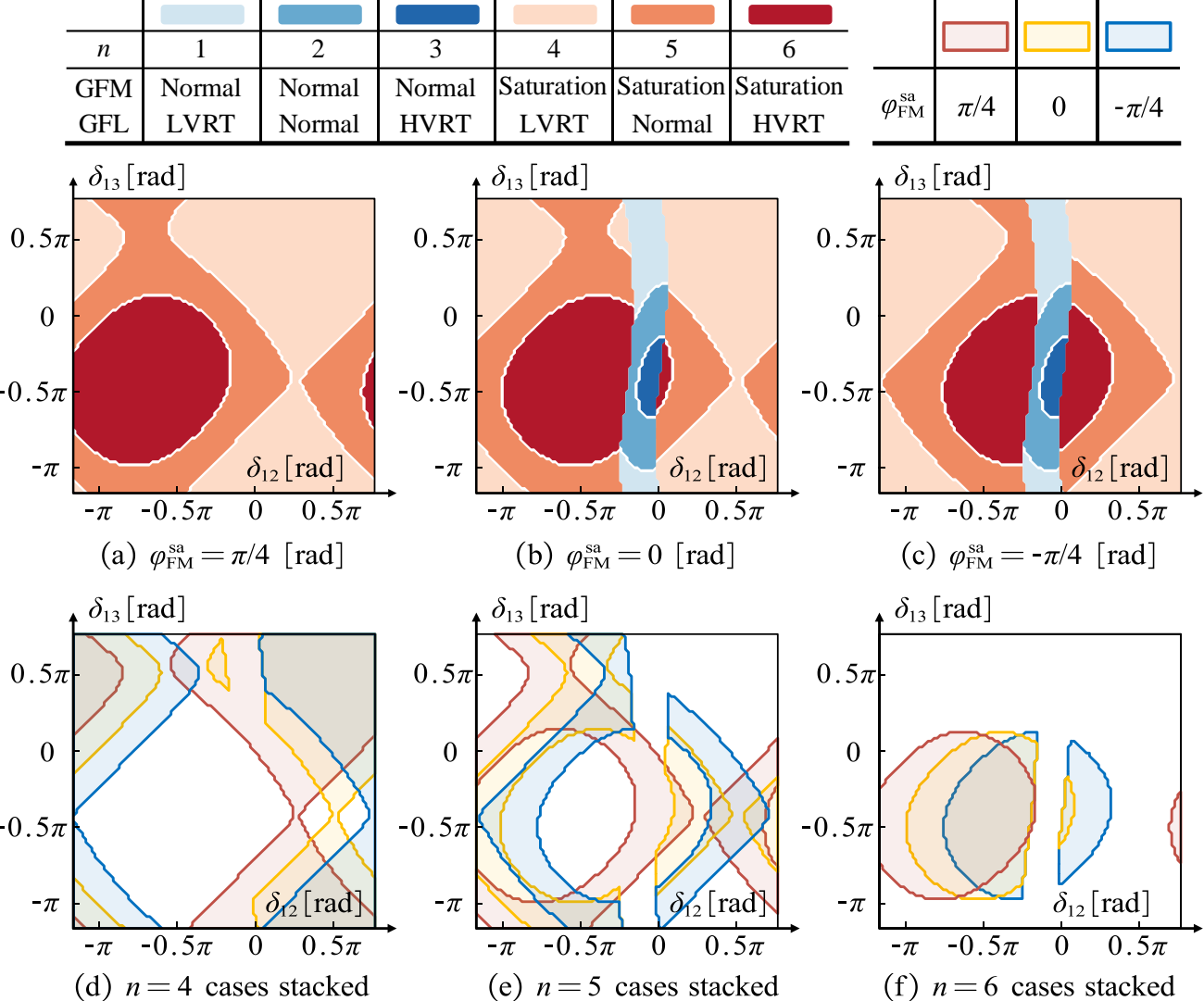

Fig. 17. Influence of current reference phase $\varphi_{FM}^{sa}$ on control combination regions during fault recovery, where (a)-(c) indicate control combination regions for $\varphi_{FM}^{sa}$ = $\pi/4$, 0, and $-\pi/4$ rad, respectively, and (d)-(f) show the stacked plots of control combination regions $n$ = 4, 5, and 6 corresponding to (a)-(c).

Fig. 18 illustrates the influence of the maximum current of GFM-RES on control combination regions. Subplots (a)-(c) correspond to cases where the ratio of the maximum current to the steady-state current increases from 1.5 to 2.5, while subplots (d)-(i) depict the evolution of the six control combination regions. It can be observed that an increase in $I_{max}$ leads to an expansion of the control combination regions with indices $n$=1~3 along the $\delta_{12}$ axis, indicating a direct enlargement of the desaturation margin. From subplots (g)-(i), it can be observed that the regions corresponding to control combinations 4~6 are compressed along the $\delta_{12}$ axis while extending toward the $\delta_{13}$ axis. This indicates that if the GFM-RES cannot effectively exit the current saturation mode, the likelihood of triggering LVRT and HVRT may increase to some extent.

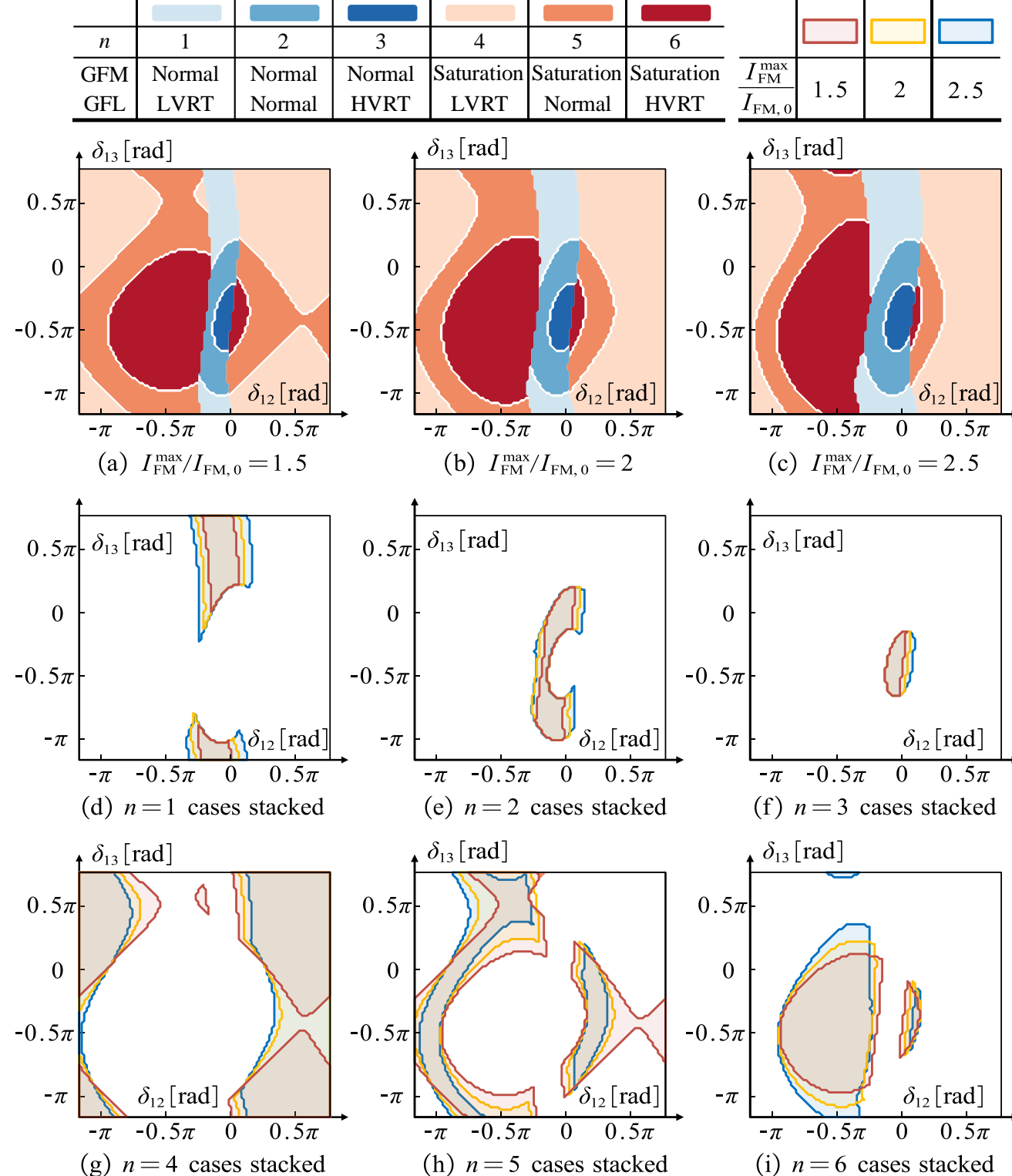

Fig. 18. Influence of max current ratio $I_{FM}^{max}/I_{FM,0}$ on control combination regions during fault recovery, where (a)-(c) indicate control combination regions for $I_{FM}^{max}/I_{FM,0}$ = 1.5, 2, and 2.5, respectively, and (d)-(i) show the stacked plots of control combination regions $n$ = 1~6 corresponding to (a)-(c).

## REFERENCES

[1] Y. Gu, T. C. Green, "Power system stability with a high penetration of inverter-based resources," *Proc. IEEE.*, vol. 111, no. 7, pp. 832-853, Jul. 2022.

[2] Y. Li, Y. Gu and T. C. Green, "Revisiting Grid-Forming and Grid-Following Inverters: A Duality Theory," *IEEE Trans. Power Syst.*, vol. 37, no. 6, pp. 4541-4554, Nov. 2022.

[3] Q. Hu, L. Fu, F. Ma, and F. Ji, "Large Signal Synchronizing Instability of PLL-Based VSC Connected to Weak AC Grid," *IEEE Trans. Power Syst.*, vol. 34, no. 4, pp. 3220-3229, Jul. 2019.

[4] Joint NERC and Texas RE Staff, "2021 Odessa disturbance," NERC, Atlanta, GA, USA, Dec. 2022.

[5] Australian Energy Market Operator, "Preliminary report - Black system event in south Australia on 28 September 2016," AEMO, Australia, Oct. 2016.

[6] National Energy System Operator, "Technical report on the events of 9", National Grid ESO, United Kingdom, Aug. 2019.

[7] L. Huang, L. Zhang, H. Xin, Z. Wang, and D. Gan, "Current limiting leads to virtual power angle synchronous instability of droop-controlled converters," in *Proc. IEEE PES Gener. Meeting*, Boston, MA, USA, Jul. 2016, pp: 1-5.

[8] L. Huang, H. Xin, Z. Wang, L. Zhang, K. Wu, and J. Hu, "Transient stability analysis and control design of droop-controlled voltage source converters considering current limitation," *IEEE Trans. Smart Grid.*, vol. 10, no. 1, pp. 578-591, Jan. 2019.

[9] K. Zhuang, H. Xin, P. Hu, and Z. Wang, "Current saturation analysis and anti-windup control design of grid-forming voltage source converter," *IEEE Trans. Energy Convers.*, vol. 37, no. 4, pp. 2790-2802, Dec. 2022.

[10] Y. Li, Y. Lu, J. Yang, X. Yuan, R. Yang, S. Yang, H. Ye, and Z. Du, "Transient stability of power synchronization loop based grid forming

converter," *IEEE Trans. Energy Convers.*, vol. 38, no. 4, pp. 2843-2859, Dec. 2023.

[11] G. Wang, L. Fu, Q. Hu, C. Liu, and Y. Ma, "Transient synchronization stability of grid-forming converter during grid fault considering transient switched operation mode," *IEEE Trans. Sustain. Energy.*, vol. 14, no. 3, pp. 1504-1515, Jul. 2023.

[12] Y. Liu, H. Geng, M. Huang, and X. Zha, "Dynamic current limiting of grid-forming converters for transient synchronization stability enhancement," *IEEE Trans. Ind. Appl.*, vol. 60, no. 2, pp. 2238 - 2248, Mar. 2024.

[13] E. Rokrok, T. Qoria, A. Bruyere, B. Francois, and X. Guillaud, "Transient stability assessment and enhancement of grid-forming converters embedding current reference saturation as current limiting strategy," *IEEE Trans. Power Syst.*, vol. 37, no. 2, pp. 1519-1531, Mar. 2022.

[14] X. Lyu, W. Du, S. M. Mohiuddin, S. P. Nandanoori and M. Elizondo, "Criteria for Grid-Forming Inverters Transitioning Between Current Limiting Mode and Normal Operation," *IEEE Trans. Power Syst.*, vol. 39, no. 4, pp. 6107-6110, Jul. 2024.

[15] X. He, H. Geng, R. Li and B. C. Pal, "Transient Stability Analysis and Enhancement of Renewable Energy Conversion System During LVRT," *IEEE Trans. Sustain. Energy.*, vol. 11, no. 3, pp. 1612-1623, Jul. 2020.

[16] X. Wang, H. Wu, X. Wang, L. Dall, and J. Kwon, "Transient Stability Analysis of Grid-Following VSCs Considering Voltage-Dependent Current Injection During Fault Ride-Through," *IEEE Trans. Energy Convers.*, vol. 37, no. 4, pp. 2749-2760, Dec. 2022.

[17] Q. Hu, L. Fu, F. Ma, G. Wang, C. Liu and Y. Ma, "Impact of LVRT Control on Transient Synchronizing Stability of PLL-Based Wind Turbine Converter Connected to High Impedance AC Grid," *IEEE Trans. Power Syst.*, vol. 38, no. 6, pp. 5445-5458, Nov. 2023.

[18] Q. Lai, C. Shen, and D. Li, "Dynamic modeling and stability analysis for repeated LVRT process of wind turbine based on switched system theory," *IEEE Trans. Power Syst.*, vol. 40, no. 3, pp. 2711-2723, May. 2025.

[19] L. Wu, W. Zhao, M. Xu, P. Xu, F. Li, Y. Yang, and Y. Pan. "Mechanism analysis and suppression of repeated voltage fluctuation considering fault ride through characteristics of the wind turbine," *J. Global Energy Interconn.*, vol. 5, no. 3, pp. 290-297, May. 2022.

[20] T. Lan, D. Sun, S. Xu, and B. Zhao, "Mechanism analysis of power system transient voltage instability dominated by low voltage ride through of renewables," *Autom. Electr. Power Syst.*, vol. 48, no. 12, pp. 58-67, Feb. 2024.

[21] S. Chen, J. Yao, Y. Liu, J. Pei, S. Huang and Z. Chen, "Coupling Mechanism Analysis and Transient Stability Assessment for Multiparalleled Wind Farms During LVRT," *IEEE Trans. Sustain. Energy.*, vol. 12, no. 4, pp. 2132-2145, Oct. 2021.

[22] Z. Tian, X. Li, X. Zha, Y. Tang, P. Sun, and M. Huang, "Transient Synchronization Stability of an Islanded AC Microgrid Considering Interactions Between Grid-Forming and Grid-Following Converters," *IEEE J. Emerg. Sel. Top. Power Electron*, vol. 11, no. 4, pp. 4463-4476, Aug. 2023.

[23] L. Karunaratne, N. R. Chaudhuri, A. Yogarathnam and M. Yue, "Nonlinear Backstepping Control of Grid-Forming Converters in Presence of Grid-Following Converters and Synchronous Generators," *IEEE Trans. Power Syst.*, vol. 39, no. 1, pp. 1948-1964, Jan. 2024.

[24] F. Sun, R. Zeng, R. Diao, B. Zhou, W. Zheng and K. Wang, "On Transient Stability Analysis of Parallel GFM Converter Systems Considering Current Saturation," *Proc. IEEE PES Gener. Meeting*., Seattle, WA, USA, 2024, pp. 1-5.

[25] W. Tang, W. Zheng, B. Zhou, Z. Yang and Y. Zhang, "Transient Stability Analysis Between Grid-Following DFIG-WT and Grid-Forming Converter in Electromechanical Timescale," *IEEE Trans. Energy Convers.*, doi: 10.1109/TEC.2024.3504860.

[26] F. Sun, R. Diao, R. Zeng, Z. Liu, B. Zhou, J. Li, and W. Tang, "Transient Stability Analysis for Grid Following Converters in Low-Inertia Power Systems by Direct Method," *Int. J. Electr. Power Energy Syst*., vol. 172, p. 111155, Nov. 2025.

[27] D. Liberzon, *Switching in systems and control*, Boston: Birkhauser, 2003.

[28] R. Zeng, R. Diao, F. Sun, B. Zhou and J. Li, "Dynamic Evolution and Stability Analysis of GFM-Based Renewable Energy Resources Considering Repeated & Continuous Current Saturation," *IEEE Trans. Sustain. Energy*., early access, doi: 10.1109/TSTE.2025.3588818.

[29] N. Baeckeland, D. Chatterjee, M. Lu, B. Johnson, and G.-S. Seo, "Overcurrent Limiting in Grid-Forming Inverters: A Comprehensive Review and Discussion," *IEEE Trans. Power Electron*., vol. 39, no. 11, pp. 14493 - 14517, Nov. 2024.

## BIOGRAPHIES

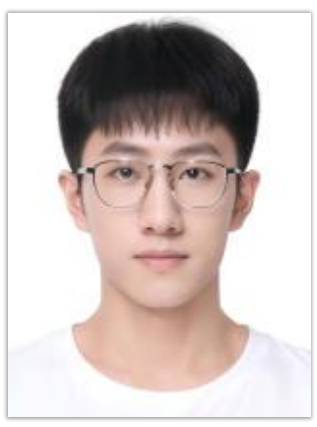

**Ruiyuan Zeng** (S'23) received his B.S. degree from Hangzhou Dianzi University, Hangzhou, China, in 2023. He is currently pursuing the M.S. degree in Electrical Engineering at the ZJU-UIUC Institute, Zhejiang University, China. His research interests include stability analysis and control of power system with high renewable energy penetration.

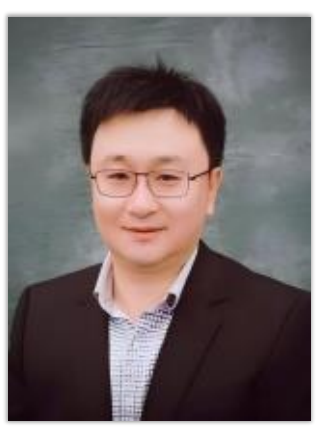

**Ruisheng Diao** (M'09-SM'15, FIET, P.E.) obtained his Ph.D. degree in Electrical Engineering at Arizona State University in 2009. He is now with the ZJU-UIUC Institute as Director, Renewable Power System Simulation & Intelligent Control and a tenured associate professor. He served as a Deputy Program Manager and Team Lead at the U.S. Department of Energy Pacific Northwest National Laboratory and as a Deputy Department Head at the Global Energy Interconnection Research Institute North America. Prof. Diao is an IET Fellow, and has served as an associate editor for IEEE Transactions on Power Systems, IEEE ACCESS, and IET Generation, Transmission & Distribution, IEEE Data Descriptions. He is also the recipient of the 2018 R&D 100 Awards and multiple IEEE PES Best Paper Awards. His research interests include high-fidelity simulation techniques and AI-based methods for grid planning and operation.

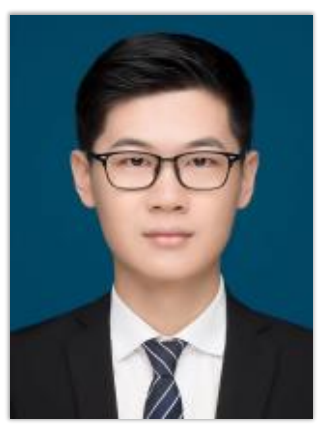

**Fangyuan Sun** (S'22) received his B.S. and M.S. degrees in Electrical Engineering from the School of Electrical Automation and Information Engineering, Tianjin University, China in 2019 and 2022, respectively. He is currently pursuing his Ph.D. degree in Electrical Engineering at Zhejiang University.

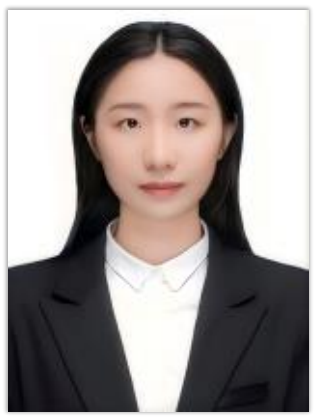

**Wangqianyun Tang** received the B.Eng. and Ph.D degrees from the School of Electrical and Electronic Engineering, Huazhong University of Science and Technology, Wuhan, China, in July 2014 and July 2020, respectively. She was a post-doctor in Electric Power Research Institute, China Southern Power Grid Co., Ltd., from 2020 to 2022. She is currently working as a senior engineer as well as researcher in SEPRI. She also works as an expert in IEC SC8A JWG5. Her current research interests include modeling and control of renewable energy, stability analysis of power-electronized power system.

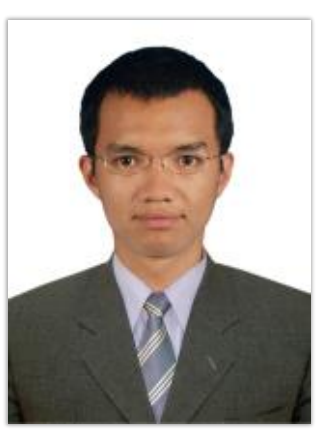

**Junjie Li** received his B.E. and the M.S. degrees in Electrical Engineering from Tsinghua University, Beijing, China, in 2009 and 2011, respectively. He is currently with the State Key Laboratory of HVDC, Electric Power Research Institute, China Southern Power Grid Co., Ltd., Guangzhou, China.

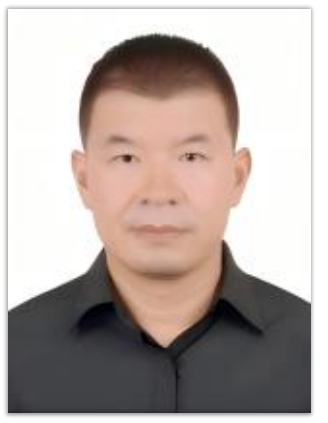

**Baorong Zhou** received his B.E. degree in Electrical Engineering from Wuhan University, Wuhan, China, in 1996, and the M.S. and Ph.D. degrees in Electrical Engineering from Tianjin University, Tianjin, China, in 2001 and 2004, respectively. He is currently with the State Key Laboratory of HVDC, Electric Power Research Institute, China Southern Power Grid Co., Ltd., Guangzhou, China.